\documentclass[pra,twocolumn,aps,superscriptaddress,longbibliography]{revtex4-1}

\usepackage{hyperref}
\usepackage{graphicx}
\usepackage{amsmath}
\usepackage{amsfonts}
\usepackage{amssymb}
\usepackage{epsfig}
\usepackage{lineno}
\usepackage{subfigure}
\usepackage{mathtools}
\usepackage{braket}
\usepackage[usenames,dvipsnames]{color}
\usepackage{setspace}
\usepackage{bm}
\usepackage{times}
\hypersetup{
      colorlinks=true,
      citecolor=blue,
      linkcolor=blue,
      urlcolor=blue}

\begin{document}
\title{Supersolid phase of the extended Bose-Hubbard model with 
       an artificial gauge field}
\author{K. Suthar}
\affiliation{Physical Research Laboratory,
             Ahmedabad - 380009, Gujarat,
             India}
\affiliation{Instytut Fizyki imienia Mariana Smoluchowskiego, \\ 
             Uniwersytet Jagiello\'nski, ulica \L{}ojasiewicza 11, 
             30-348 Krak\'ow, Poland}
\author{Hrushikesh Sable}
\affiliation{Physical Research Laboratory,
             Ahmedabad - 380009, Gujarat,
             India}
\affiliation{Indian Institute of Technology Gandhinagar,
             Palaj, Gandhinagar - 382355, Gujarat,
             India}
\author{Rukmani Bai}
\affiliation{Physical Research Laboratory,
             Ahmedabad - 380009, Gujarat,
             India}
\affiliation{Indian Institute of Technology Gandhinagar,
             Palaj, Gandhinagar - 382355, Gujarat,
             India}
\author{Soumik Bandyopadhyay}
\affiliation{Physical Research Laboratory,
             Ahmedabad - 380009, Gujarat,
             India}
\affiliation{Indian Institute of Technology Gandhinagar,
             Palaj, Gandhinagar - 382355, Gujarat,
             India}
\author{Sukla Pal}
\affiliation{Physical Research Laboratory,
             Ahmedabad - 380009, Gujarat,
             India}
\affiliation{Department of Physics, Centre for Quantum Science, and Dodd-Walls
             Centre for \\ Photonic and Quantum Technologies, University of
             Otago, Dunedin, New Zealand}
\author{D. Angom}
\affiliation{Physical Research Laboratory,
             Ahmedabad - 380009, Gujarat,
             India}
\date{\today}

\begin{abstract}

 We examine the zero and finite temperature phase diagrams of 
soft-core bosons of the extended Bose-Hubbard model on a square optical 
lattice. To study various quantum phases and their transitions we employ 
single-site and cluster Gutzwiller mean-field theory. We have observed that 
the Mott insulator phase vanishes above a critical value of nearest-neighbour 
interaction and the supersolid phase occupies a larger region in the phase 
diagram. We show that the presence of artificial gauge field enlarges the 
domain of supersolid phase. The finite temperature destroys the crystalline 
structure of the supersolid phase and thereby favours normal fluid to 
superfluid phase transition. The presence of an envelope harmonic potential 
demonstrates coexistence of different phases and at 
$z~k_{B}T\geqslant V$, thermal energy comparable 
and higher to the long-range interaction energy, the supersolidity of the 
system is destroyed. 
\end{abstract}

\maketitle

\section{Introduction} 
 Ultracold atomic systems have played an important role in the study of 
quantum many-body systems. In particular, the novel experimental developments 
in manipulating ultracold atoms in optical lattices have led to the 
realization of new quantum states and quantum phase transitions in strongly 
correlated systems~\cite{lewenstein_07,bloch_08}. In recent years, there has 
been a surge of interest in understanding the supersolid (SS) phase which is 
characterized by the simultaneous appearance of a crystalline and an 
off-diagonal long-range orders~\cite{leggett_70,chester_70}. This phase breaks 
two continuous symmetries: the phase invariance of the superfluid (SF) and 
translational invariance to form crystal. Although the SS phase was predicted 
in liquid $^{4}$He a long time ago~\cite{andreev_69,thouless_69}, the 
experimental observation of supersolidity in liquid $^{4}$He remains 
elusive~\cite{kim_04,kim_04a,kim_12}. However, the quest for SS phase has 
gained new impetus following the remarkable theoretical insights and 
experimental achievements in ultracold atoms in optical lattices, which are 
excellent quantum many-body systems to observe SS phase as these are clean and 
controllable. Recently, the characteristic signature of SS phase has been 
observed in ultracold 
atoms~\cite{leonard_17,li_17,tanzi_19,bottcher_19,chomaz_19} and this phase 
may emerge by tuning the bosonic interactions of different length 
scales~\cite{landig_16}. Furthermore, the excitation
spectrum and various properties of SS phase have been observed in recent 
quantum gas experiments~\cite{natale_19,tanzi_19a,guo_19}.

On the theoretical aspects, the existence of SS phase has been studied 
using the extended Bose-Hubbard model (eBHM). The {\em checkerboard} 
supersolid phase of hard-core bosons is thermodynamically unstable towards 
phase separation and this phase is not stabilized by next NN (NNN) 
interaction~\cite{batrouni_00}. However, in the soft-core model, due to 
larger Fock-space and possibility of higher number fluctuations, the SS 
order is stable with NN interaction~\cite{otterlo_95,sengupta_05,ohgoe_12b}. 
Similarly, SS phase emerges in the honeycomb lattices when the hard-core limit 
of bosons is relaxed~\cite{wessel_07,gan_07}. The existence and stability of 
supersolidity have also been confirmed for bosons with infinite-range 
cavity-mediated interactions~\cite{flottat_17}. The SS phase has been 
explored in various lattice systems, such as one-dimensional (1D) 
chain~\cite{mathey_09,batrouni_06} and ladder~\cite{sachdeva_17}, 
two-dimensional (2D) 
square~\cite{goral_02,sengupta_05,scarola_05,yi_07,ng_08,capogrosso_10,
bandyopadhyay_19},
triangular~\cite{wessel_05, heidarian_05, melko_05,boninsegni_05,sen_08,
yamamoto_12}, honeycomb~\cite{wessel_07,gan_07}, 
kagome~\cite{isakov_06,huerga_16}, bilayer lattice of dipolar 
bosons~\cite{trefzger_09}, and three-dimensional (3D) cubic 
lattice~\cite{yi_07,yamamoto_09,xi_11,ohgoe_12}. The eBHM with artificial 
gauge field has been studied to examine the fractional quantum 
Hall~\cite{kuno_17} and vortex-solid states~\cite{kuno_15}.

In this work we investigate theoretically the presence of SS phase of 
soft-core bosons in 2D square optical lattice with long-range 
interaction and artificial gauge field. The long-range interaction can be 
realized with the dipolar ultracold atoms~\cite{pasquiou_11, baier_16}. And, 
it is possible to introduce artificial gauge field with 
lasers~\cite{aidelsburger_11,miyake_13,atala_14,kennedy_15}. 
For our studies we use the single-site and cluster mean-field theories.
We show that the combined effect of the long-range interaction and artificial 
gauge field increases the domain of the SS phase. In particular, we examine 
the effect of magnetic flux quanta on the SS-SF phase boundary in the presence 
of the NN interaction. Furthermore, to relate with experimental realizations, 
we incorporate the effects of thermal fluctuations arising from finite 
temperatures.

The paper is organized as follows. In Sec.~\ref{theory_gw} we introduce the 
model considered in our study and describe theoretical approach employed. Here 
we provide description of the single-site, cluster and finite temperature
Gutzwiller (GW) mean-field theories. The ground-state phase diagrams and study 
of dipolar atoms in the confining potential are presented in 
Sec.~\ref{results}. Finally, we conclude with the key findings of the present 
work in Sec.~\ref{conc}.

\section{Theoretical Methods}
\label{theory_gw}


\subsection{Extended Bose-Hubbard model}
Consider a system of bosonic atoms with long-range interactions in a 2D square 
optical lattice in the presence of synthetic magnetic field. The temperature 
of the system is low such that all the atoms occupy the lowest Bloch band. 
Such a system is well described by eBHM, and the Hamiltonian of the model is
\begin{eqnarray}
  \hat{H}_{\rm eBHM} &= &-\sum_{p,q}\left(J_x 
                         \hat{b}^{\dagger}_{p+1,q}\hat{b}_{p,q} 
                         + J_y \hat{b}^{\dagger}_{p,q+1}\hat{b}_{p,q} 
                         + \rm{H.c.} \right) \nonumber\\
                      && +\sum_{p,q} \hat{n}_{p,q} 
                         \left[ \left(\epsilon_{p,q} - \mu \right) 
                         + \frac{U}{2} (\hat{n}_{p,q} - 1) \right] \nonumber\\
                      && + \sum_{\langle \xi \xi' \rangle} 
                        V_{\xi,\xi'}~\hat{n}_{\xi}\hat{n}_{\xi'},
\label{bh_ham}
\end{eqnarray}      
where $p(q)$ is the lattice site index along $x(y)$ direction, 
$\hat{b}^{\dagger}_{p,q}$ ($\hat{b}_{p,q}$) is the bosonic operator which 
creates (annihilates) an atom at the lattice site $(p,q)$, $\hat{n}_{p,q} = 
\hat{b}^{\dagger}_{p,q} \hat{b}_{p,q}$ is the boson number operator, $J_x$ and 
$J_y$ are the tunneling or hopping strength between two NN sites along $x$ and
$y$ directions, respectively, $\epsilon_{p,q}$ is the offset energy arising 
due to the presence of external envelope potential, $\mu$ is the chemical 
potential, and $U>0$ is the on-site interatomic interaction. Here $\xi$ is a
combination of lattice indices in 2D, that is, $\xi \equiv (p,q)$ and 
$\xi' \equiv (p',q')$ are neighbouring sites to $\xi$. The long-range 
interaction $V_{\xi,\xi'}$ is given by
\begin{equation}
  V_{\xi,\xi'} =\begin{cases}
    V_1 & \text{ if $|\mathbf{r}_{pq}-\mathbf{r}_{p'q'}|=a$}, \\
    V_2 & \text{ if $|\mathbf{r}_{pq}-\mathbf{r}_{p'q'}| =\sqrt{2}a$}, \\
    0   & \text { otherwise}, 
  \end{cases}
\label{v1v2}
\end{equation}
where $a$ is the lattice spacing, ${\bf r}_{pq} = (pa, qa)$ is the lattice 
site coordinates. The parameters $V_1\geqslant 0$ and $V_2\geqslant 0$ are the 
NN and NNN interactions, respectively. 
In the present work, we consider $V_2/V_1 = 1/2\sqrt{2}$ which corresponds to
inverse cube power law of isotropic dipolar interaction. These terms encapsulate
the observable effects of the dipolar interaction in the system. The higher
$V_1$ as compared to $V_2$ tends to induce checkerboard density pattern for
density wave (DW) and SS phases. This minimal model captures the essential
physics arising due to the hopping induced competition among different solid
orders and the superfluidity. We also consider this model to examine the 
ground states of inhomogeneous dipolar bosons at zero and finite temperatures 
in optical lattices with an envelope confining harmonic potential.


\subsection{Artificial gauge field}
The long-range interaction in the above Hamiltonian, Eq.~(\ref{bh_ham}), is 
characteristic or inherent to the internal state of the atomic species. In 
terms of the many-body physics, the nature of the correlation can further be 
modified through the introduction of artificial gauge field. The presence of 
the artificial gauge field modifies the Hamiltonian to
\begin{eqnarray}
  \hat{H}_{\rm eBHM} &=&-\sum_{p,q}\left(J_x e^{i 2\pi\alpha q}
                         \hat{b}^{\dagger}_{p+1,q}\hat{b}_{p,q} 
                        + J_y \hat{b}^{\dagger}_{p,q+1}\hat{b}_{p,q} 
                        + \rm{H.c.} \right) \nonumber\\
                      &&+\sum_{p,q} \hat{n}_{p,q} 
                         \left[ \left(\epsilon_{p,q} - \mu \right) 
                        + \frac{U}{2} (\hat{n}_{p,q} - 1) \right] \nonumber\\
                      &&+ \sum_{\langle \xi \xi' \rangle} 
                        V_{\xi,\xi'}~\hat{n}_{\xi}\hat{n}_{\xi'},
\label{bh_ham_gauge}
\end{eqnarray} 
where the strength of the magnetic field is reflected in the number of flux 
quanta per plaquette $\alpha = (e/\hbar) \int d{\mathbf r}.\mathbf{A(r)}$. 
Here, $0\leqslant\alpha<1$, and $\mathbf{A(r)}$ is the vector potential which 
gives rise to synthetic magnetic field $\mathbf{B = \nabla\times A}$. In the 
presence of the synthetic magnetic field, atoms acquire a $2\pi\alpha$ phase 
when they hop around a plaquette. This results into a phase shift in the 
hopping strength of the model. Physically, the synthetic magnetic field 
introduces a force on the atoms which is equivalent of the Lorentz force on a 
charged particle in the presence of external magnetic field. The system is 
then a charge neutral analogue of the quantum Hall system in condensed matter 
systems. For the present study, we consider Landau gauge, where the vector 
potential $\mathbf{A(r)} = -By\hat{x}$. Hence, for the homogeneous system, at 
zero magnetic field the system possesses the translational invariance along 
both axes, whereas in the presence of magnetic field the system preserves the 
invariance only along the $x$-axis of the lattice.


\subsection{Gutzwiller mean-field theory}
To study the ground-states of the systems described by the model Hamiltonians 
in Eq.~(\ref{bh_ham}) and (\ref{bh_ham_gauge}) and their properties we use 
single-site Gutzwiller mean-field (SGMF) and cluster Gutzwiller mean-field 
(CGMF) theories. The later is the extension of SGMF which incorporates the 
correlation within a cluster of neighbouring sites exactly. In the SGMF 
theory~\cite{rokshar_91,krauth_92,sheshadri_93, rukmani_18, pal_19, 
bandyopadhyay_19}, the bosonic operators are 
expanded about their expectation values as
\begin{subequations}
  \begin{eqnarray}
    \hat{b}_{p,q} = \phi_{p,q} + \delta\hat{b}_{p,q},\\
    \hat{b}^{\dagger}_{p,q} = \phi^{*}_{p,q} + \delta\hat{b}^{\dagger}_{p,q}.
  \end{eqnarray}
\end{subequations}
Therefore, the product of the creation and annihilation operators which occurs
in the hopping term can be written as 
\begin{equation}
  \hat{b}^{\dagger}_{p,q} \hat{b}_{p',q'} \approx 
  \phi^{*}_{p,q}\hat{b}_{p',q'} 
  + \hat{b}^{\dagger}_{p,q} \phi_{p',q'} - \phi^{*}_{p,q}\phi_{p',q'}, 
\end{equation}
where second order terms in the fluctuation $\delta\hat{b}_{p,q}$ are 
neglected. Here, $\phi_{p,q} = \langle \hat{b}_{p,q} \rangle$ is the SF order 
parameter of the system. Using above approximation in the Hamiltonian, 
Eq.~(\ref{bh_ham_gauge}), the single-site mean-field Hamiltonian is
\begin{eqnarray}
  \hat{H}^{\rm MF}_{p,q}&=&- \bigg[J_x e^{i 2\pi\alpha q}
                               \left(\phi^{*}_{p+1,q}\hat{b}_{p,q} 
                               - \phi^{*}_{p+1,q}\phi_{p,q}\right) 
                               \nonumber\\
                         & &+ J_y\left(\phi^{*}_{p,q+1}\hat{b}_{p,q}
                               - \phi^{*}_{p,q+1}\phi_{p,q}\right) 
                               + \rm{H.c.} \bigg]
                               \nonumber \\
                         & &+ \left[(\epsilon_{p,q} - \mu) 
                            +  \frac{U}{2} (\hat{n}_{p,q}-1)
                               \right] \hat{n}_{p,q} 
                               \nonumber \\
                         & &+ \sum_{\langle \xi,\xi' \rangle} 
			       V_{\xi,\xi'}~\left(\hat{n}_{\xi} 
			       \langle\hat{n}_{\xi'}\rangle
			     - \langle\hat{n}_{\xi}\rangle
			       \langle\hat{n}_{\xi'}\rangle\right)
\label{ss_ham}
\end{eqnarray}
and the total Hamiltonian of the system is 
\begin{equation}
  \hat{H}_{\rm MF} = \sum_{p,q} \hat{H}^{\rm MF}_{p,q}. 
\end{equation}
Here, the neighbouring lattice sites are coupled through $\phi_{p,q}$, the SF 
order parameter. And therefore the eigenstate of the entire lattice is the 
product of single-site states. Accordingly, the many-body wave function of 
the ground state of the system is given by the Gutzwiller ansatz
\begin{equation}
  \ket{\Psi} = \prod_{p,q}\ket{\psi}_{p,q} = \prod_{p,q} 
                                             \left(\sum^{N_b}_{n = 0} 
                                        c^{(p,q)}_{n} \ket{n}_{p,q}\right), 
\label{gw_ansatz}
\end{equation}
where $\ket{\psi}_{p,q}$ is the single-site ground state, $N_b$ is the number 
of occupation basis or maximum number of bosons at each lattice site, 
$\ket{n}_{p,q}$ is the occupation or Fock state of $n$ bosons occupying the 
site $(p,q)$ and $c^{(p,q)}_{n}$ are the coefficients 
of the occupation state. The normalization of the wave function leads to the 
normalization of $c^{(p,q)}_{n}$ at each lattice site as 
$\Sigma_{n} |c^{(p,q)}_{n}|^2 = 1$. 
Using the above ansatz the SF order parameter 
$\phi_{p,q} = \bra{\Psi} \hat{b}_{p,q} \ket{\Psi}$ is obtained as 
\begin{equation}
  \phi_{p,q} = \sum^{N_b}_{n=0} \sqrt{n}~c^{(p,q)*}_{n-1} c^{(p,q)}_{n}.
\label{phi} 
\end{equation}
Similarly, the occupancy of each lattice site
$n_{p,q} = 
\bra{\Psi}\hat{b}^{\dagger}_{p,q}\hat{b}_{p,q}\ket{\Psi}$ is
\begin{equation}
  n_{p,q} = \sum^{N_b}_{n=0} n |c^{(p,q)}_n|^2.
\label{expn} 
\end{equation}
The two parameters $\phi_{p,q}$ and $n_{p,q}$, together serve to define the 
quantum phases of the system. 
Using the mean-field Hamiltonian, Eq.~(\ref{ss_ham}), the 
total energy of the system $E=\bra{\Psi}\hat{H}_{\rm MF}\ket{\Psi}$ is
obtained as a sum of the single-site energies 
$E_{p,q}= \bra{\Psi}\hat{H}_{p,q}^{\rm MF}\ket{\Psi}$. 
And,  $E$ is minimized self consistently with the Eqs.~(\ref{phi}) 
and~(\ref{expn}) to obtain the ground state of the system.


 In the CGMF theory, a lattice of dimension $K\times L$ is partitioned into 
$W$ clusters of size $M\times N$, that is $W = (K\times L)/(M\times N)$
~\cite{buonsante_04,yamamoto_09a,pisarski_11,mcintosh_12,luhmann_13,rukmani_18,
pal_19}. Then, the hopping terms of the model are decomposed into two types. 
One is the exact term which corresponds to hopping within the cluster, and the 
other is the inter-cluster hopping between lattice sites which lie on the 
boundary of two neighbouring clusters. The latter is defined by coupling 
through the mean-field or the SF order parameter. The Hamiltonian of a cluster 
is
\begin{eqnarray}
  \hat{H}_C  &=& - {\sum_{p,q \in C}}' \left(J_x e^{i 2\pi\alpha q}
                  \hat{b}_{p+1,q}^{\dagger}\hat{b}_{p,q} 
                 +J_y \hat{b}_{p,q+1}^{\dagger}\hat{b}_{p,q}
                 +{\rm H.c.} \right)\nonumber\\
              && -\sum_{p, q\in \delta C}
                  \left(J_x {\rm e}^{i 2\pi\alpha q} 
                  (\phi^c_{p+1,q})^{\ast}\hat{b}_{p,q} 
                 +J_y (\phi^c_{p,q+1})^{\ast}\hat{b}_{p,q} 
                 + {\rm H.c.} \right) \nonumber \\
              && +\sum_{p,q \in C} \left[(\epsilon_{p,q} - \mu) \hat{n}_{p,q}
                 +\frac{U}{2}\hat{n}_{p,q}(\hat{n}_{p,q}-1)\right] \nonumber\\
              && +\sum_{\langle\xi\xi'\rangle \in C} 
		  V_{\xi,\xi'}~\hat{n}_{\xi}~\hat{n}_{\xi'} 
                 +\sum_{\langle\xi\xi'\rangle \in \delta C}
		  V_{\xi,\xi'}~\hat{n}_{\xi}~\langle\hat{n}_{\xi'}\rangle,
\label{cgmf_ham}         
\end{eqnarray}
where the model parameters $J_x$, $J_y$, $U$ and $V_{\xi,\xi'}$ are defined 
as in SGMF and prime in the first summation indicates that the $(p+1,q)$ and 
$(p,q+1)$ lattice sites are also within the cluster. Here, $\delta C$ in the 
second summation represents the lattice sites at the boundary of the clusters 
and $(\phi^c_{p,q}) = \sum_{p',q' \not\in C} \langle \hat{b}_{p',q'}\rangle$
is the SF order parameter at the lattice site which lies at the boundary of 
neighbouring cluster. As hopping parameter, the long-range interaction term 
also has two contributions, one is within the cluster which is exact, and the 
other is inter-cluster interaction at the boundary which is defined through 
the mean occupancy $\langle \hat{n}_{\xi'} \rangle$. The matrix elements of
$\hat{H}_C$ are, then, calculated in terms of the cluster basis states 
\begin{equation}
  \ket{\Phi_c}_\ell = \prod_{q=0}^{N-1}\prod_{p=0}^{M-1} \ket{n_p^q},
\end{equation}
where $\ket{n_p^q}$ is the occupation number basis at the $(p,q)$ lattice 
site, and $\ell \equiv \{n_0^0, n_1^0, \ldots, n_{M-1}^0, n_0^1, n_1^1,\ldots
n_{M-1}^1, \ldots, n_{M-1}^{N-1}\}$ is the index quantum number to identify 
the cluster state. After diagonalizing the Hamiltonian, we can get the ground 
state of the cluster as
\begin{equation}
  |\Psi_c\rangle = \sum_{\ell} C_\ell\ket{\Phi_c}_\ell,
\end{equation}
where $C_\ell$ are components of the eigenvector, and naturally satisfy the 
normalization condition $\sum_{\ell}|C_\ell|^2 = 1$. The ground state of the 
entire $K\times L$ lattice, as in SGMF, is the direct product of the cluster 
ground states
\begin{equation}
  \ket{\Psi^c_{\rm GW}} = \prod_k\ket{\Psi_c}_k,
\label{cgw_state}
\end{equation}
where $k$ is the cluster index and varies from 1 to 
$W=(K\times L)/(M\times N)$. The SF order parameter $\phi$, as in 
Eq.~(\ref{phi}), can be computed in terms of the cluster states. The
average occupancy of the $k$th cluster can also be computed similarly.


\subsection{Finite temperature Gutzwiller mean-field theory}
\label{gw_finit_T}
At finite temperature, the thermal fluctuations modify the properties of the 
system, and observable properties are the thermal averages. To calculate the 
thermal averages we need the entire eigenspectrum. So, in the SGMF, we retain 
the entire energy spectrum $E^l_{p,q}$ and the eigenstates 
$\ket{\psi}^l_{p,q}$ obtained from the diagonalization of the single-site 
Hamiltonian $\hat{H}^{\rm MF}_{p,q}$ in Eq.~(\ref{ss_ham}). Then, we evaluate 
the single-site partition function of the system
\begin{equation}
  Z = \sum_{l=1}^{N_b}e^{-\beta E^l},
\end{equation}   
where $\beta = 1/k_{B}T$ and $T$ is the temperature of the system. At finite 
$T$, the region in the phase diagram with $\phi = 0$ and the real 
occupancy $\langle \hat{n}_{p,q} \rangle$ is identified as the normal fluid 
(NF) phase. Similarly, in the CGMF, the partition function is defined in terms
of all the eigenvalues $E^{l}_k$ and eigenfunctions $\ket{\Psi_c}^{l}_k$ of 
each $k$th cluster from all the $W$ clusters. 

From the definition of the partition function, in the SGMF, the thermal 
average of $\phi_{p,q}$ is
\begin{equation}
	\langle \phi_{p,q}\rangle = \frac{1}{Z}\sum_{i=0}^{N_b}
					       \prescript{i}{p,q}{\bra{\psi}}
                              \hat{b}_{p,q} e^{-\beta E^i} \ket{\psi}^i_{p,q},
\end{equation}
where $\langle\ldots\rangle$ represents the thermal averaging. Similarly,
the occupancy or the density at finite $T$ is defined as 
\begin{equation}
  \langle\langle \hat{n}_{p,q} \rangle\rangle = \frac{1}{Z}\sum_{i=0}^{N_b}
                                             \prescript{i}{p,q}{\bra{\psi}}
			   \hat{n}_{p,q} e^{-\beta E^{i}}\ket{\psi}^i_{p,q}.
\end{equation}
The average occupancy is $\langle n \rangle 
= \sum_{p,q}\langle\langle \hat{n}_{p,q} \rangle\rangle /(K\times L)$. These 
definitions can be extended to the CGMF by replacing the single-site states 
and energies with that of the cluster.


\subsection{Characterization of phases}
\label{char_phase}
 The ground state phases in eBHM and the phase boundaries are 
characterized by several order parameters. At zero temperature, the ground 
states of eBHM support two incompressible and two compressible phases. The
incompressible phases are Mott insulator (MI) and DW and compressible phases are
SS and SF. Among these phases, DW and SS are due to the long-range interaction
between the atoms. The characteristic distinction between the DW and MI is 
their density distributions : the MI has commensurate occupancy whereas the DW 
phase has incommensurate occupancy with long-range crystalline order. 
The insulating phases have zero $\phi$ and integer occupancy of each site 
$n_{p,q}$, on the other hand, the compressible phases have finite $\phi$ and
real $n_{p,q}$. We identify the phase boundaries between the MI(DW) and SF(SS) 
phase based on the order parameter $\phi$ and $n_{p,q}$. The SS phase has 
long-range crystalline order in $\phi$ and $n_{p,q}$ whereas SF phase has 
uniform density distribution of atoms. The DW and SS phase are better described
in terms of the sublattices $A$ and $B$ such that the NN sites belong to 
different sublattices. The relative average occupancy $\langle \Delta n \rangle$
is another order parameter which can distinguish the DW and SS phases from MI 
and SF phases. In the SGMF method, for a $K\times L$ lattice it is defined as
\begin{equation}
  \langle \Delta n \rangle = \frac{1}{K\times L}~{\sum_{\langle\xi\xi'\rangle}} 
                             |\langle \hat{n}_{\xi} \rangle 
                            - \langle \hat{n}_{\xi'} \rangle|,
\end{equation}
where $\langle \hat{n}_{\xi} \rangle \equiv \langle \hat{n}_{A} \rangle$ and 
$\langle \hat{n}_{\xi'} \rangle \equiv \langle \hat{n}_{B} \rangle$ are 
sublattice occupancies. The similar expression can be defined for CGMF theory. 
For the DW and SS phases $\langle \Delta n \rangle$ is nonzero, and in 
particular, it is integer and real for the DW and SS phases, respectively. But,
for MI and SF phases $\langle \Delta n \rangle$ is zero as 
$\langle \hat{n}_{p,q} \rangle$ is uniform. Table~\ref{tab_phase} summarizes 
the classification of all the phases discussed in the present work.
\begin{table}[ht]
  \begin{tabular}{l | c | c | c | c }
   \hline \hline
   Quantum phases       &${n}_{p,q}$ & $\phi_{p,q}$ & $\langle\Delta n\rangle$ 
	                                            & $\kappa$ \\
   \hline 
   Mott insulator (MI)  & integer & $0$      & $0$                & $0$ \\
   Density wave (DW)    & integer & $0$      & $\neq 0$ (integer) & $0$ \\
   Supersolid (SS)      & real    & $\neq 0$ & $\neq 0$ (real)    & $\neq 0$ \\
   Superfluid (SF)      & real    & $\neq 0$ & $0$                & $\neq 0$ \\ 
   Normal fluid (NF)    & real    & $0$      & $0$ or real   & $\neq 0$ \\
   \hline
  \end{tabular}
  \caption{Classification of phases at zero and finite temperatures.}
  \label{tab_phase} 
\end{table}

At finite temperature, a NF phase present in the system which is distinguish
from incompressible MI and DW phases by examining the local density variance
which is also the measure of the local compressibility~\cite{mahmud_11,parny_12}
\begin{equation}
  \kappa = \frac{\partial\langle \hat{n} \rangle}{\partial\mu}
	 = \beta\left(\langle \hat{n}^2_{p,q} \rangle 
	    - {\langle \hat{n}_{p,q} \rangle}^2 \right).
\end{equation}
This quantity defines the density fluctuations of the system. $\kappa$ is zero 
for MI and DW phases whereas it is nonzero for the NF phase. We use these 
order parameters to obtain the phase boundaries between various phases at zero
and finite temperatures. In the phase diagrams, discussed in next section, the 
incompressible phases are indicated by their sublattice occupancies 
$(n_{A}, n_{B})$ with $n_{A} = n_{B}$ for MI and $n_{A}\neq n_{B}$ for DW 
phase.


\section{Results and discussions}
\label{results}
  The standard BHM shows two phases, the incompressible MI 
phase corresponding to commensurate integer filling, and the compressible SF 
phase which has finite $\phi$. The SF-MI quantum phase transition was 
observed by tuning the depth of the optical lattice~\cite{greiner_02}. 
In eBHM, the introduction of the NN interaction changes the phase diagram 
through the emergence of two more phases. First is the DW, which sandwiches 
the MI lobes at low values of NN interaction, and second is the SS phase, it 
occurs as envelope around the DW lobes. In this work, we first examine the 
phase diagram for the homogeneous systems. We, then, study the impact of 
artificial gauge field on the phase diagram by considering $\alpha = 1/2$. 
For comparison with experimental realizations, we also study with envelope 
potential.  
\begin{figure}[ht]
 {\includegraphics[width=8.5cm]{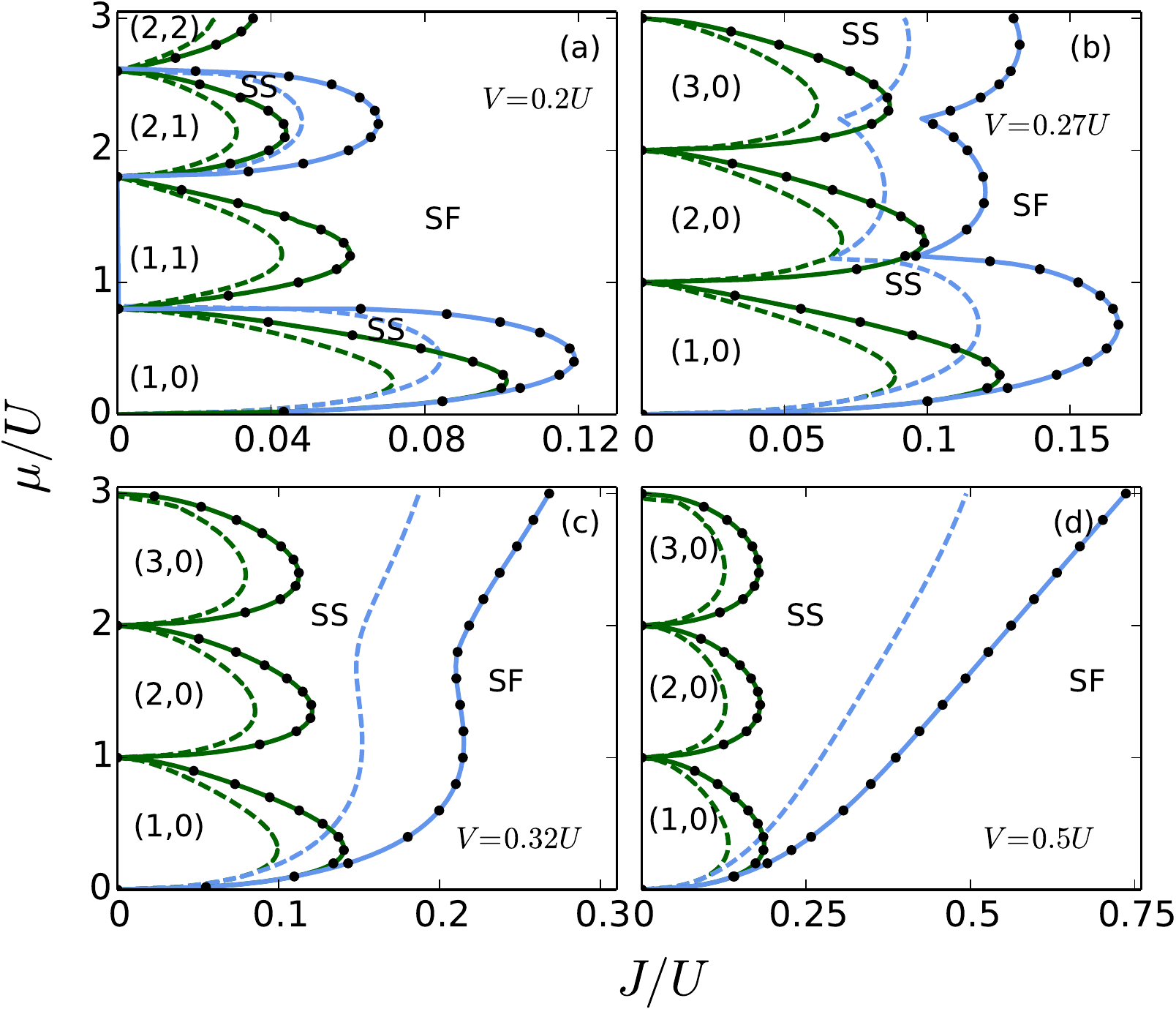}}
   \caption{The phase diagrams of the eBHM with uniform hopping amplitude
	  ($J_x = J_y = J$) for $V/U = 0.2, 0.27, 0.32$, and $0.5$ (a-d). 
	  The plots show the phase boundaries for two cases $\alpha=0$ and 
	  $\alpha=1/2$. The dashed (solid) green line indicates the 
	  MI-SF, DW-SF and DW-SS phase boundaries for $\alpha=0$ ($\alpha=1/2$)
	  case. The dashed (solid) blue line represents the SS-SF phase 
	  boundary for $\alpha=0$ ($\alpha=1/2$) case. The phase boundaries 
	  for $\alpha=1/2$ case are obtained using the Landau gauge. The 
	  data points obtained using the symmetric gauge are represented by
	  black solid filled circles. Here, the DW and MI phases are indicated 
	  by their sublattice occupancies $(n_A,n_B)$.}
\label{ph_diag}
\end{figure}


\subsection{Homogeneous case}
\label{homo_case}
\begin{figure}[ht]
 {\includegraphics[width=8.5cm]{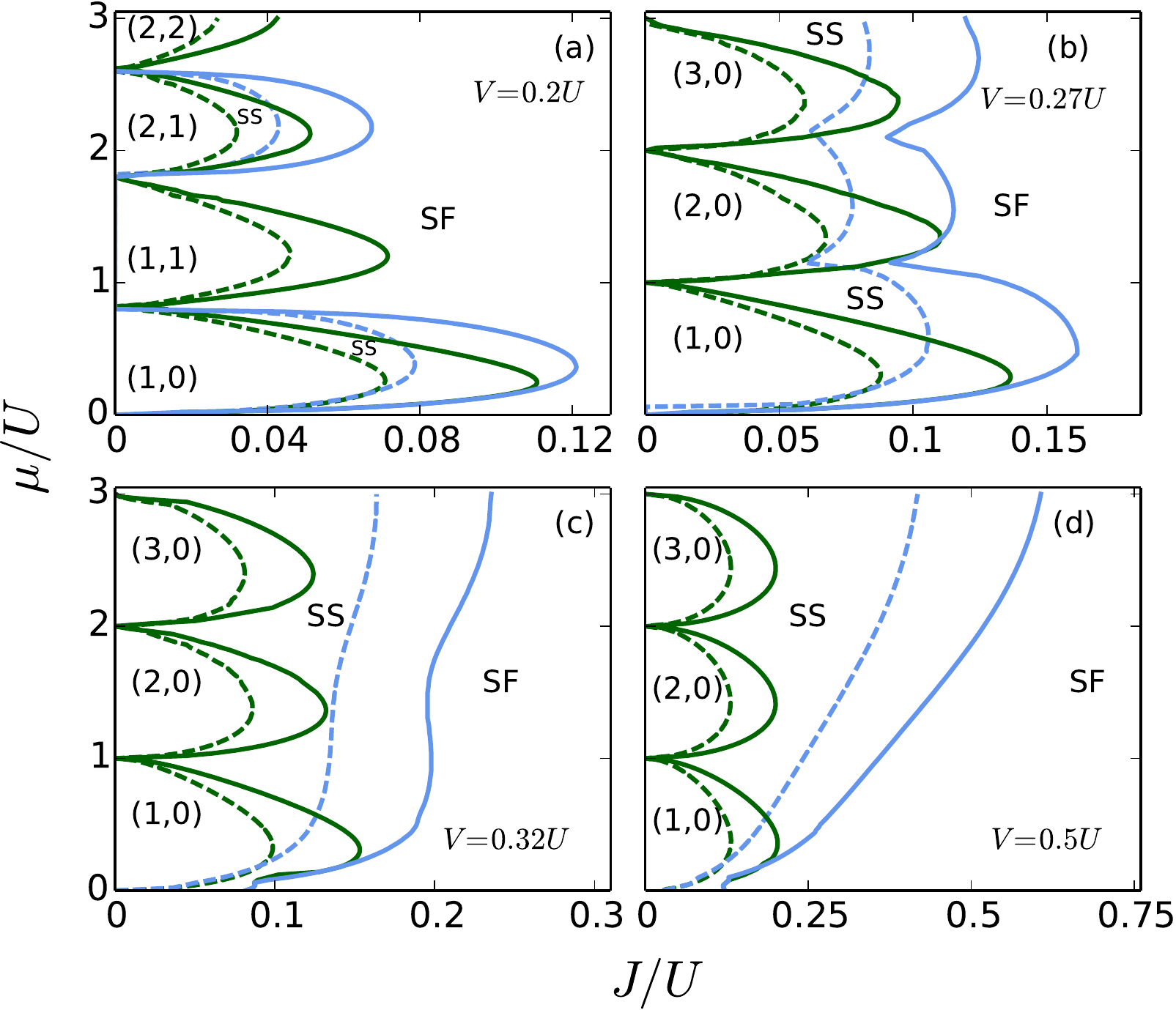}}
 \caption{The phase diagrams of the eBHM obtained from the CGMF theory with 
	  $2\times 2$ clusters for uniform hopping amplitude ($J_x = J_y = J$) 
	  at $V/U = 0.2, 0.27, 0.32$, and $0.5$ (a-d). The plots show the 
	  phase boundaries for two cases $\alpha=0$ and $\alpha=1/2$. 
	  The left dashed (solid) green line indicates the MI-SF, DW-SF and 
          DW-SS phase boundaries for $\alpha=0$ ($\alpha=1/2$) case. The 
          right dashed (solid) blue line represents the SS-SF phase boundary 
          for $\alpha=0$ ($\alpha=1/2$) case. Here the DW and MI phases are 
          indicated by their 
	  sublattice occupancies $(n_A,n_B)$.}
\label{ph_diag_2by2}
\end{figure}

 The phase diagram of the eBHM obtained from the SGMF is as shown in 
Fig.~\ref{ph_diag} for different values of $V$. It is important 
to note that here $V$ is NN interaction which is $V_1$ of long-range 
interaction [Eq.~(\ref{v1v2})] and $V_2 = 0$. In the phase diagram, the 
incompressible DW and MI phases are identified by their sublattice occupancies 
$(n_A,n_B)$. We observed that for $zV<U$, where $z$ is the coordination 
number of the system, the ground state alternates between MI and DW phases, 
and regions of SS phase occur as envelopes around the DW phase lobes. On 
increasing $V$, at a critical value $zV_c = U$ the MI lobes are transformed 
into DW phase. For $V\geqslant V_c$, the SS phase occupies a larger region 
in the phase diagram. As $V$ is increased, the other observable effect is 
the critical value of the hopping strength $J_c/U$ for the DW-SS transition 
also increases. At higher values, when $zV\gtrsim 1.5 U$, the SS-SF phase 
boundary is like a linear function of the $J$, and this is discernible from 
Fig.~\ref{ph_diag}(d). In particular, the phase boundary is linear when 
$zJ/U > 1$. These findings are in good agreement with the previous work of 
Iskin~\cite{iskin_11}. The numerical results of the phase boundaries are in 
good agreement with the analytical predictions of mean-field decoupling theory
~\cite{iskin_12}. For example, the critical hopping $J_c/U$ for DW(1,0) to 
SS transition at $V/U=0.32$ is $0.0994$ and analytical theory predicts the same 
value for these set of parameters. Similarly, we find the MI-SF phase 
boundaries from our study are consistent with the mean-field decoupling 
theory~\cite{iskin_12}. To examine the importance of the inter-site correlation 
effects the phase diagram using $2\times2$ CGMF method is as shown in 
Fig.~\ref{ph_diag_2by2}. It is qualitatively similar to the one based on SGMF 
in Fig.~\ref{ph_diag}. But, there are several quantitative differences. 
First, the MI phase lobe is enhanced whereas DW phase lobe is suppressed. As 
an example, for $V/U = 0.2$ the tip of the DW$(1,0)$ lobe is at 
$J_c/U = 0.0717$ using SGMF theory, but with CGMF theory it is decreased 
to $0.0709$. Although, the difference between the $J_c$ using both theories
is very small but using higher clusters CGMF one can get significant
difference of $J_c$'s. This is apparent from the cluster finite-size scaling of 
the cluster sizes discussed in the next subsection. Second, at higher values 
of $V$ and $\alpha=1/2$, the SS-SF phase boundary commences at $J/U\approx 0$ 
and $\mu/U\approx 0$ with the SGMF theory. On the other hand, with the CGMF 
theory the SS-SF boundary starts at finite value of $J/U$ and $\mu/U=0$ 
[Fig.~\ref{ph_diag_2by2}(c,d)]. Third, compared to the SGMF results, with the 
CGMF theory we obtain SS domains which are smaller in size. This is due to 
the better representation of fluctuations in the CGMF theory. 
Our results demonstrate the greater accuracy of the CGMF theory 
by correcting the overestimation of the SS domain obtained from
the single-site mean-field theory. This observation is also consistent with 
similar comparison between results obtained from the single-site mean-field 
theory and quantum Monte Carlo (QMC)~\cite{ohgoe_12b,flottat_17}. And, 
finally, for higher values of $V/U$ the SS-SF boundary is linear at higher 
$\mu$ with SGMF theory. But, it is curved with the CGMF theory. 
Qualitatively, the value of $J_c$ obtained using 
SGMF and CGMF are close to the QMC results available in the literature.  
The value of $J_c/U$ of the DW(1,0) - SS quantum phase transition at 
$zV/U=1$ obtained using SGMF and CGMF methods are $0.0841$ and $0.0832$, 
respectively, and these values are close to QMC result of 
$0.0822$~\cite{ohgoe_12b}. Furthermore, we also carried out additional 
computations to compare the $J_c/U$ of DW-SF quantum phase transition at the 
half-filling with the QMC predictions~\cite{batrouni_95}. For example, at 
$V/U\approx 0.35$, using SGMF we obtain $J_c/U=0.175$. Using CGMF theory this 
value  is improved to $0.153$. And, this is in very good agreement with QMC 
result of $0.143$ reported in Ref.~\cite{batrouni_95}. Further improvements 
in the values of the $J_c$ of various quantum phase transitions is possible 
when clusters of larger sizes are considered. This is also evident from the 
cluster finite-size scaling analysis discussed in next subsection.

As mentioned earlier, to study the effect of artificial gauge field, we choose
$\alpha = 1/2$. So, hereafter {\em with artificial gauge field} we mean 
$\alpha=1/2$. And, {\em without artificial gauge field} means $\alpha=0$. 
The artificial gauge field modifies the phase boundaries of MI, DW and SS 
phases, and the changes are discernible from the phase diagrams in 
Fig.~\ref{ph_diag}. For example, for the $(1,0)$ DW lobe at $V/U = 0.2$, 
the tip of the lobe is enhanced from $J_c/U\approx 0.0717$ by $40\%$ to $0.101$ 
with artificial gauge field. This enhancement in the insulating 
lobe and the phase boundaries of DW(MI) - SS(SF) transition with artificial 
gauge field are in agreement with the mean-field decoupling 
theory~\cite{iskin_12}. For example, the $J_c/U$ of DW(1,0)-SS
quantum phase transition with $\alpha = 1/2$ at $V/U = 0.32$ is $0.1406$ 
which is same as the value obtained from the analytical 
theory~\cite{iskin_12}. The tip of the SS lobe is enhanced from 
$J_c/U\approx 0.084$ to $\approx 0.119$, and these changes together implies 
a larger domain of SS phase surrounding the DW(1,0) phase. These changes 
arise from the localizing effect of the Landau quantization, associated with 
the artificial gauge field, on the itinerant bosons. In addition, there are 
major differences between the SGMF and CGMF phase diagrams. For example, the 
tip of the $(1,0)$ DW lobe is increased from $J_c/U\approx 0.101$ in SGMF 
to $J_c/U =  0.110$ in CGMF. This implies that the effect of correlation 
due to finite magnetic flux is better captured by the CGMF method.
We perform the stability analysis of SS phase by computing the average 
occupancy $\langle n \rangle$ as a function of $\mu$ ~\cite{sengupta_05, 
flottat_17}. The unstable phase is characterized by a discontinuity in 
$\langle n \rangle$ as $\mu$ is varied. In our study, we fix the $J/U$ and 
vary $\mu/U$ such that the SS phase is traversed, and then compute the 
average occupancy. As an example, at $J/U = 0.12$ and $V/U = 0.32$ for 
$\alpha = 0$ case, we do not observe any discontinuity in $\langle n \rangle$
as a function of $\mu/U$. This confirms the stability of the SS phase of 
the soft-core bosons. We further include the NNN interaction 
term and find that the SS phase remains stable. The stability 
of soft-core SS phase is consistent with the QMC results of 
Ref.~\cite{sengupta_05, ohgoe_12b}. In addition, our analysis also 
demonstrates the stability of SS phase in the presence of the artificial gauge 
field.

 To check the gauge-invariance of the phase boundaries, 
we also compute the phase boundaries for $\alpha \neq 0$ using SGMF and 
CGMF with the symmetric gauge. For the symmetric gauge, the vector potential 
$\mathbf{A(r)} = (1/2)(-By\hat{x} + Bx\hat{y})$. We observe that the phase 
boundaries obtained from the symmetric gauge are in good agreement with the
results from Landau gauge. This can be seen from Fig.(\ref{ph_diag}), where 
the filled black circles are the phase boundaries obtained using
the symmetric gauge. For example, with the symmetric gauge, the tip of 
DW(1,0) lobe for $V/U = 0.2$ is $J_c/U = 0.101$ and which is identical
with the Landau gauge result. Similarly, for the same value of $V/U$, the
CGMF results with $2\times2$ is $J_c/U=0.110$. And, this result is invariant
under Landau and symmetric gauges. The SS-SF phase boundaries are also 
gauge-invariant as well. To illustrate, consider the SS-SF phase transition 
for $V/U = 0.2$. With $\mu/U = 0.44$, both the symmetric and Landau gauges 
give $J_c/U = 0.119$. And, in the case of CGMF, the value $J_c/U = 0.1207$ 
obtained with $2\times2$ clusters is gauge invariant. Thus, the SGMF and 
CGMF methods give gauge-invariant phase boundaries for incompressible to 
compressible and SS-SF quantum phase transitions. This is consistent with 
the general principle that the observable quantities are gauge 
invariant~\cite{boada_10,moller_10}. And, it shows that the numerical 
methods we have used are robust as the results are gauge invariant.

\begin{figure}[ht]
 {\includegraphics[width=7.8cm] {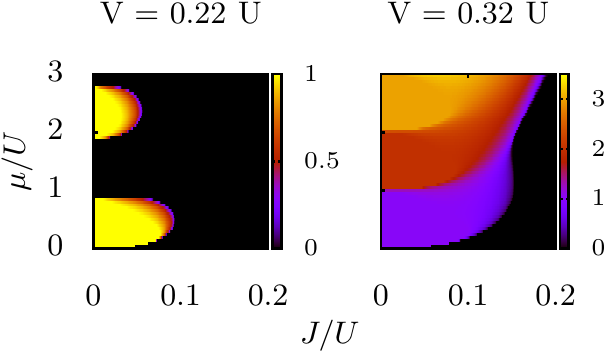}} 
 \caption{The relative average occupancy of two consecutive sites are shown
          as a function of the chemical potential $\mu$ and hopping parameter 
          $J$. The NN interaction is shown at the top of the figures.}
\label{avg_rel_num}         
\end{figure}

 The nature of the DW-SS transition is better represented by 
$\langle\Delta n \rangle$ and values for $V = 0.22 U$ and $0.32 U$ 
corresponding to $V < V_c$ and $V > V_c$ are shown in Fig.~\ref{avg_rel_num}. 
In the figure, the dark regions correspond to MI and SF phases and the region 
in other colors correspond to DW and SS phases. For $V = 0.22 U$, the regions 
in yellow color are DW phases and regions in other shades correspond to SS. 
The gradient in the shades indicates that the transition from DW to SS in 
terms of $\langle \Delta n \rangle$ is smooth. For the case of $V = 0.32 U$, 
there are no dark regions in the neighbourhood of $J/U\approx0$. This is due 
to the absence of MI lobes, and is consistent with the phase diagram shown in 
Fig.~\ref{ph_diag} as all the MI lobes are transformed to DW lobes. The nature 
of the DW phases are apparent and visible from color gradient in 
Fig.~\ref{avg_rel_num} as the colors indicate the difference in the occupancy 
of two neighbouring lattice sites. As in the case of $V = 0.22 U$, regions 
with a color gradient indicate the SS phase and overall the relative average 
occupancy is in agreement with the phase diagram in Fig.~\ref{ph_diag}. 
However, the phase diagram in terms of $\langle \Delta n \rangle$ provides a 
richer descriptions of the two phases, DW and SS, unique to the eBHM vis-a-vis
BHM. And, the appropriate order parameter to examine the regions of SS phase. 

\begin{figure}[ht]
 {\includegraphics[width=8.5cm] {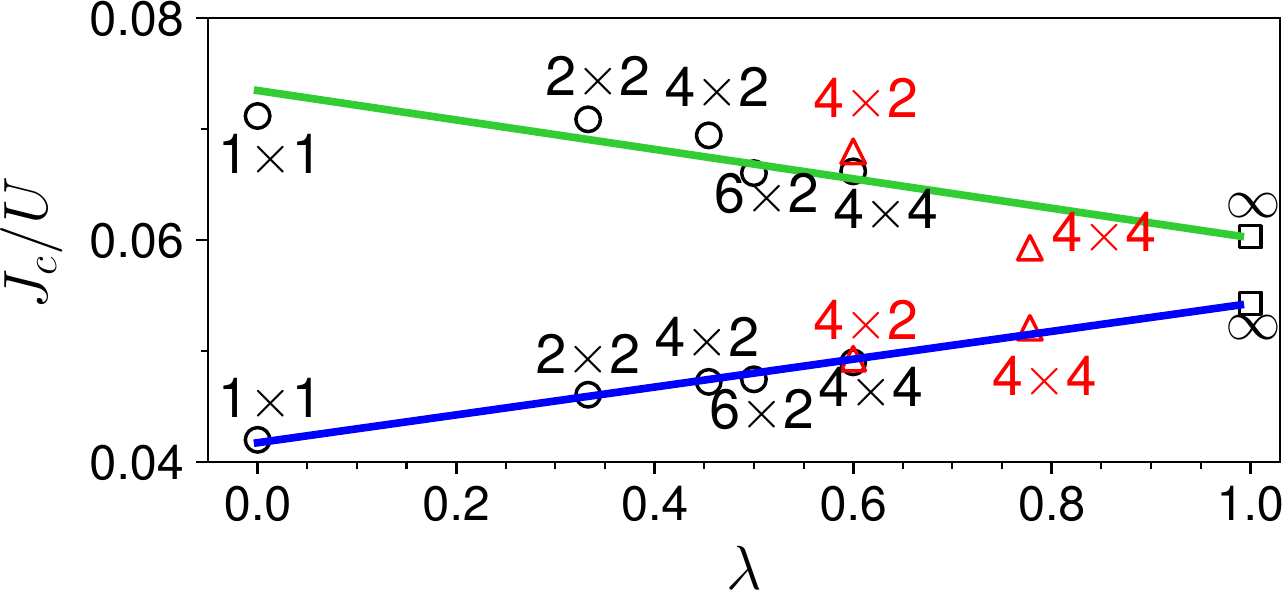}}
 \caption{The cluster finite-size scaling of $J_c$ for (1,1) MI-SF (blue) and 
	  (1,0) DW-SS (green) transition for different cluster sizes. The 
	  circles (triangles) represent the critical values obtained using 
	  periodic boundary conditions along both (one) spatial dimension. The 
	  square symbols represent the exact values in the thermodynamic 
	  limit. The scaling is performed for $\alpha = 0$ and $V=0.2U$.}
\label{fscal}
\end{figure}


\subsection{Cluster finite-size scaling}
 The results obtained from the SGMF and CGMF do provide qualitatively
correct phase diagrams. And, this can further be improved using cluster finite-size
scaling analysis. Such an analysis provides the location of the phase
boundary in the thermodynamic limit. As a case study, we examine the location
of the DW and MI lobe tips $J_c$ for $\alpha = 0$, and $V=0.2U$.
To implement the finite scaling analysis, we use a series of square and 
rectangular clusters $N_{\rm C} = 2\times 2$, $4\times 2$, $6\times 2$, and 
$4\times 4$. In addition, we also include the results of $4\times 2$ and 
$4\times 4$ clusters with exact hopping along one spatial direction. Here, 
we consider clusters with even number of sites along $x$ and $y$ directions,
as only these clusters generate checkerboard order. To obtain the thermodynamic 
limit, we introduce the scaling parameter 
$\lambda = N_{\rm B}/(N_{\rm B}+N_{\rm{\delta B}})$ which varies from $0$ to 
$1$. Here, $N_{\rm B}$ is the number of bonds within the cluster and 
$N_{\rm{\delta B}}$ is the number of bonds at the boundary which couples the
cluster to it's neighbours through the mean-field term~
\cite{yamamoto_12a,luhmann_13}. The parameter $\lambda$ is a measure of the
atomic correlations taken into account by using clusters of various sizes. In 
the extreme limits, the SGMF $N_{\rm C} = 1\times 1$ and exact $N_{\rm C} = 
\infty$ results correspond to $\lambda = 0$ and $\lambda = 1$, respectively. 
Thus, the value of $J_c$ improves as $\lambda$ of the 
cluster approaches to $1$.

The cluster finite-size scaling analysis for the (1,0) DW and (1,1) MI phases with
$\alpha = 0$, and $V=0.2U$ are shown in Fig.~\ref{fscal}. As seen from the 
figure, the location of the MI lobe tip $J_c$, increases with 
the cluster size. The SGMF favours the SF phase leading to underestimation of 
the MI phase. And, CGMF with larger cluster sizes enhance the MI lobe.
Based on the linear fit the thermodynamic limit of $J_c$
is $0.05428$. One key feature is that, the results from the periodic
boundary condition with exact hopping in one direction lie closer to the 
fitted line. And, more importantly, these have higher $\lambda$ as these have
less bonds coupled to the mean field. The scaling behaviour of $J_c$ is 
in good agreement with the scaling results of MI-SF in 
BHM~\cite{mcintosh_12,luhmann_13}.

For the case of (1,0) DW phase, $J_c$  decreases with increasing cluster size.
In other words, the domain with DW order shrinks with larger clusters and
implies that SGMF theory overestimates DW order. This is corrected with the 
fluctuations incorporated with larger clusters. The trend is opposite to the
MI phase case. From the cluster finite-size scaling analysis, the linear fit gives the 
thermodynamic limit of $J_c$ as $0.06021$. This trend of $J_c$ is also 
reported in earlier studies of hard-core eBHM~\cite{yamamoto_12,yamamoto_12a}. 
Furthermore, the scaled $J_c$ values of eBHM phase boundaries at $zV/U=1$ are 
in good agreement with the QMC results~\cite{ohgoe_12b}. In the 
presence of artificial gauge field, the qualitative features of the scaling 
analysis will be modified, and, the predicted exact critical values 
of the transitions will differ from the $\alpha=0$ case.


\subsection{Finite temperature effects}
\begin{figure}[ht]
 \includegraphics[width=8.5cm] {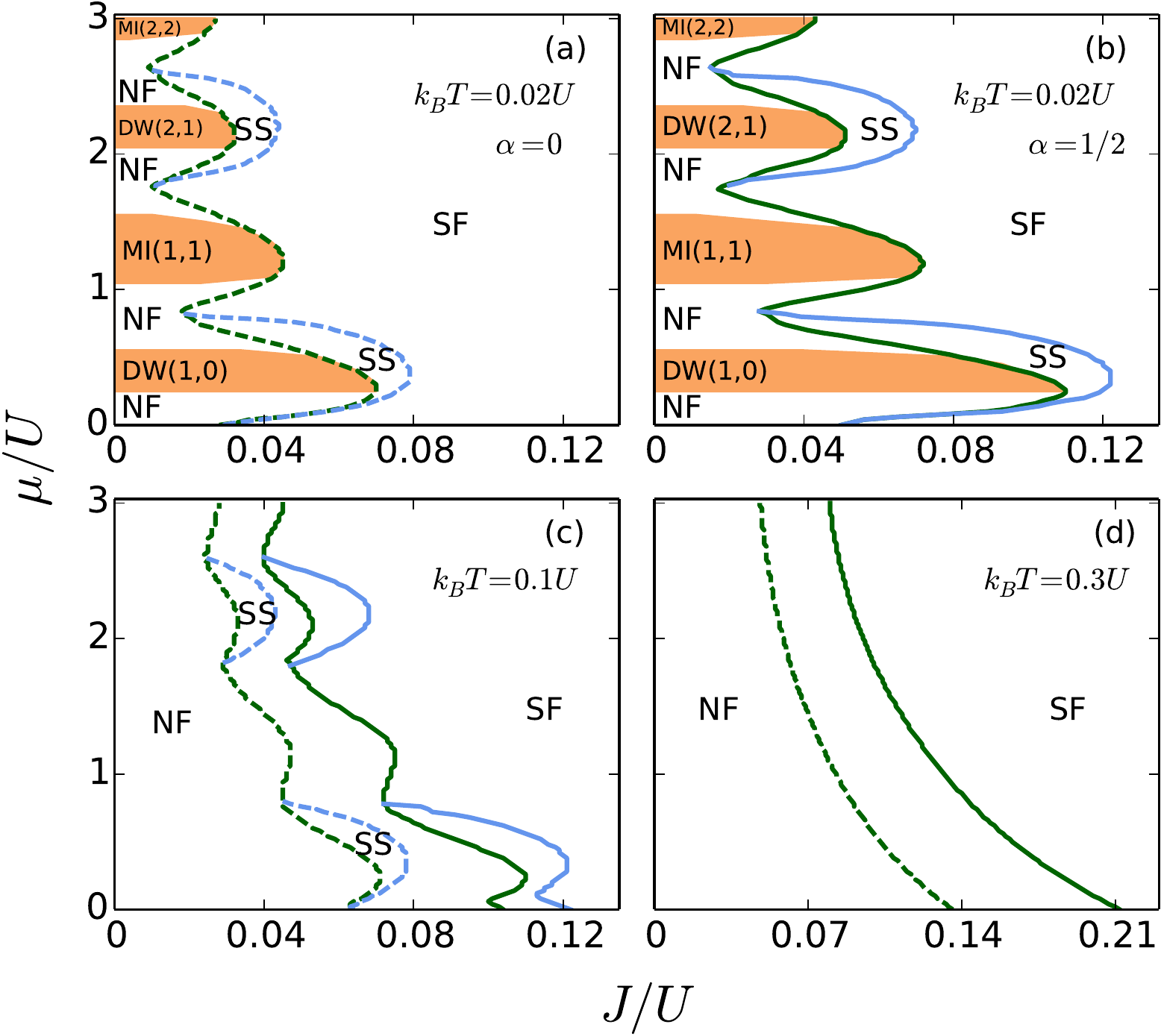}
 \caption{The finite temperature phase diagrams of eBHM obtained from CGMF 
	  theory with $2\times 2$ clusters. The orange striped region 
	  mark the DW and MI phases. (a) Phase diagram for $\alpha = 0$ at 
	  $k_{B}T = 0.02 U$. The dashed green line marks MI-SF, NF-SF, DW-SS, 
	  and NF-SS phase boundaries. And, the dashed blue line marks the 
	  SS-SF phase boundary. 
	  (b) Phase diagram for $\alpha = 1/2$ at $k_{B}T = 0.02 U$. 
	  The solid green line marks MI-SF, NF-SF, DW-SS, and NF-SS phase 
	  boundaries. And, the solid blue line marks the SS-SF phase boundary. 
	  Phase diagrams for $\alpha = 0$ and $1/2$ at (c) $k_{B}T = 0.1 U$ and 
	  (d) $k_{B}T = 0.3 U$ with the combined color scheme of (a) and (b).
	  Here, the NN interaction $V=0.2U$.}
\label{ph_diag_T_al0}
\end{figure}
 Thermal fluctuations associated with finite temperatures are an essential 
feature of experimental observations. Although the zero temperature phase 
diagrams do provide key insights and qualitative understanding, to relate with
the experimental results it is essential to incorporate thermal fluctuations.
We do this through the approach outlined in Section~\ref{gw_finit_T}. As 
mentioned earlier, the SS phase is yet to be observed in the eBHM and this 
could be due to the sensitivity of the phase to the thermal fluctuations. At 
zero temperature, SS phase appears in the system at a finite value of the NN 
interaction $V$. In Fig.~\ref{ph_diag_T_al0}, we show the finite temperature 
phase diagrams obtained using $2\times 2$ clusters in the CGMF method. As we 
have demonstrated and by others \cite{buonsante_04,yamamoto_09a,pisarski_11,
mcintosh_12,luhmann_13,rukmani_18,pal_19} that the results with CGMF are more 
reliable, hereafter we only consider the results from CGMF theory. From the 
plots in Fig.~\ref{ph_diag_T_al0}, a distinguishing feature of the thermal 
fluctuations is the emergence of the NF phase. The thermal fluctuation melts 
both the MI and DW phases and destroys the SF phase at the MI-DW or DW-DW 
boundaries. 

  To be more specific at $k_{B}T = 0.02 U$, as shown in 
Fig.~\ref{ph_diag_T_al0}(a-b), the orange stripes mark the DW and MI phases and 
the NF phase exist outside of these. The MI and NF both have zero $\phi$ and 
commensurate densities. The difference is that the MI has integer commensurate 
density, but NF has real commensurate density. The DW phase, on the other 
hand, has checkerboard density but with integer values. 
The NF region around the DW phase has incommensurate real 
density. By observing the local density variance or the compressibility, 
the incompressible DW and MI phases can be differentiated from NF phase 
as mentioned in Section~\ref{char_phase}.
On comparing the plots in Fig.~\ref{ph_diag_T_al0}(a) and 
Fig.~\ref{ph_diag_T_al0}(b), it is clear that the larger MI and DW lobes with 
finite $\alpha$ are retained at finite temperatures and hence the larger SS 
domain as well. At intermediate temperatures, both the MI and DW are entirely 
transformed into the NF phase but a portion of the SS lobes survives. This is 
visible in the phase diagram at $k_{B}T = 0.1 U$ shown in the 
Fig.~\ref{ph_diag_T_al0}(c). From the plots in the figure, the quantitative 
differences with and without the artificial gauge field is also visible. With 
artificial gauge field, the domain of the NF and SS phases are larger. For 
example, at $\mu=0$ the NF extends upto $J/U = 0.064$ and $J/U=0.104$ for zero 
and finite $\alpha$, respectively. This trend of larger extent of NF phase 
with finite $\alpha$ extends to higher values of $\mu$. Upon further increase 
in temperature the crystalline order of the SS phase is destroyed and it 
vanishes from the phase diagram. At $k_{B}T \approx 0.3 U$, as shown in 
Fig.~\ref{ph_diag_T_al0}(d), only the NF and SF phases are present in the 
system. At $\mu/U = 0$ the NF-SF phase boundary is located at $J/U=0.136$ and 
$J/U = 0.212$ for zero and finite $\alpha$, respectively. The separation 
between the location of the phase boundaries is reduced as $\mu/U$ increased. 
This is to be expected as the size of the DW lobes decrease with increasing 
$\mu/U$. As in the zero temperature case, the phase boundaries of the finite 
temperature phase diagrams are also gauge-invariant. We confirmed by 
analyzing the phase boundaries using symmetric gauge. As an illustration, 
consider the DW-SS transition at $\mu/U = 0.4$ and $k_{B}T/U = 0.02$ with 
$\alpha = 1/2$, this point is chosen from the phase diagram in 
Fig.~\ref{ph_diag_T_al0}(b). We obtain $J_c/U = 0.098$ using symmetric gauge 
and this value is in excellent agreement with the Landau gauge result. 
Similarly, for the same value of $\mu/U$ the SS-SF transition occurs at
$J_c/U=0.122$ and this value is gauge-independent. At higher temperature, 
$k_{B}T/U = 0.3$, the phase boundary of the NF-SF transition is also 
gauge-invariant. Hence, the comparison of $J_c/U$ for various quantum
phase transitions using Landau and symmetric gauges demonstrate the 
gauge-invariance of the phase boundaries obtained with the numerical methods 
we have adopted in this work.

\begin{figure}[ht]
 {\includegraphics[width=9.0cm] {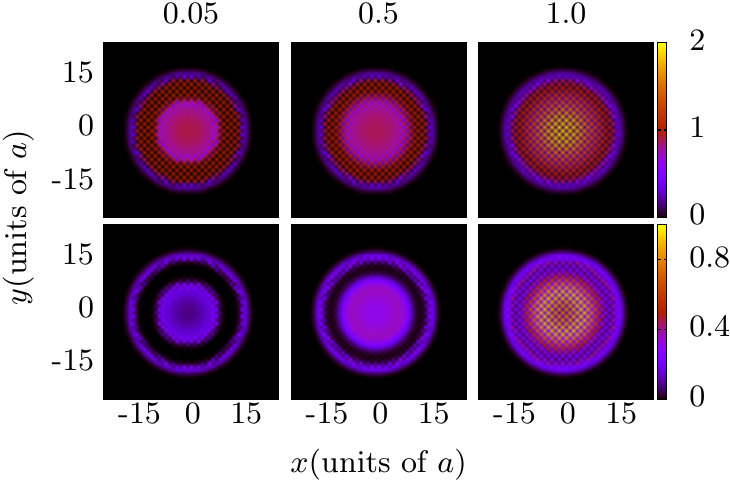}}
 \caption{The zero temperature density distribution (upper panel) and SF order 
	  parameter (lower panel) of a $50\times 50$ square lattice for 
          different values of dipolar interaction strengths $V/U$. 
	  The $V/U$ value is shown at the top of the 
          figures. Here $x$ and $y$ are in units of the lattice spacing $a$.}
\label{dip_kbt_0}
\end{figure}


\subsection{Inhomogeneous case}
 The system considered so far is uniform and we emulate it with a 
$12\times 12$ lattice with periodic boundary conditions. However, in most of 
the quantum gas experiments the optical lattice has a confining envelope 
potential. Most often the external envelope potential is a harmonic 
oscillator. Hence, the inhomogeneity arising from this confining potential is 
another factor to be considered for comparison with the experimental 
observations. Therefore, we examine the ground-state of eBHM in a 
$50\times 50$ square lattice with SGMF theory. In which the external harmonic 
potential is incorporated in the chemical potential through the offset energy 
$\epsilon_{p,q} = \Omega(p^2 + q^2)$. Here $\Omega$ is the strength of the 
confining potential. The parameters of the system considered are 
$J/V = 0.1$, $\mu/V = 2.8$ and $\Omega/V = 0.01$~\cite{trefzger_08}. 
The value of $\Omega$ is such that the atomic density outside the lattice 
potential is zero. The finite $\Omega$ modifies the local $\mu$ and therefore 
the ground state exhibits coexistence of various phases. Here $V$ is the 
strength of the dipolar interaction and the range of the potential is 
considered upto second nearest neighbours. Therefore, for the long-range 
interaction Eq.~(\ref{v1v2}), $V_{1} = V$ and $V_{2} = V/2\sqrt{2}$.

  To determine the changes in the competing phases we examine the ground state
of the system at $V/U = 0.05$, $0.5$, and $1.0$. These values cover the weak 
and strong limits of the dipolar interaction. In the experiments these regimes 
are reachable using Feshbach resonance in the dipolar atoms like 
Cr~\cite{werner_05}, Er~\cite{frisch_14} and Dy~\cite{maier_15,lucioni_18}. 
As in the previous cases, to study the effects of the thermal fluctuations
we consider three different values of $k_{B} T/V = 0$, $0.2$, and $0.3$. At 
zero temperature, the profiles of $n_{p,q}$ and $\phi_{p,q}$ corresponding to 
$V/U = 0.05$, $0.5$, and $1.0$ are shown in Fig.~\ref{dip_kbt_0}. For weak 
dipolar interaction $V/U = 0.05$ the density is nearly uniform in the central 
region. The corresponding $\phi_{p,q}$ though nearly uniform shows a dip
around the center. But, it is more uniform at the intermediate strength of the 
dipole interaction $0.5~U$. In both of these cases, $V/U = 0.05$ and $0.5$, 
the central SF region is surrounded by the $(1,0)$ DW phase and this is 
evident from ring-shaped profile of the checkerboard density pattern in the 
figure. The other key feature is that the domain of the DW phase gets 
narrowed as $V/U$ is increased and above a critical value $V/U = 0.8$ there is 
a quantum phase transition from DW phase to SS phase. This happens when both 
the interaction strengths, on-site and dipole interactions are comparable. As 
shown in Fig.~\ref{dip_kbt_0}, for $V/U = 1$, there is a large region around 
the center where both the density and SF order parameter show checkerboard 
distributions. This is the signature of the SS phase.    
\begin{figure}[ht]
 {\includegraphics[width=9.0cm] {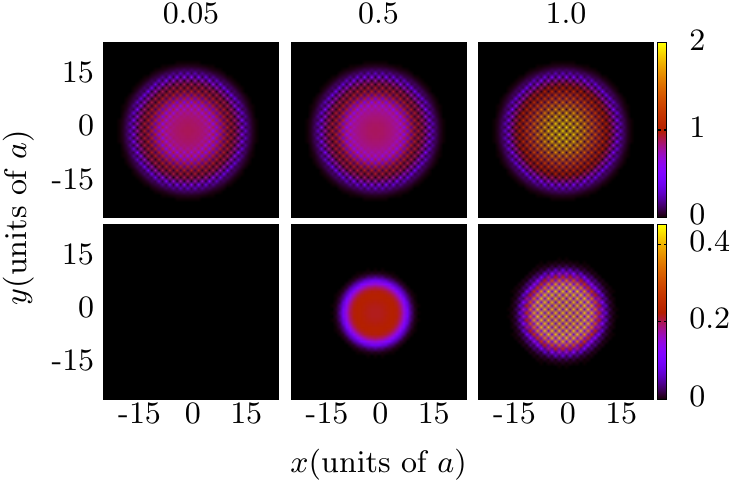}}
 \caption{The density distribution (upper panel) and SF order parameter
          (lower panel) of a $50\times 50$ square lattice at 
          $k_{B}T/V = 0.2$ for different values of dipolar strengths $V/U$. 
	  The $V/U$ value is shown at the top of the 
          figures. Here $x$ and $y$ are in units of the lattice spacing $a$.}
\label{dip_kbt_p2}
\end{figure}

 Next, to relate to the experimental realizations we incorporate the finite 
temperature effects. For weak dipolar interaction the thermal fluctuations 
leads to the melting of the SF phase. This is evident from the density and SF 
order parameter corresponding to $V/U=0.05$ at $k_{B}T/V = 0.2$ as shown in 
Fig.~\ref{dip_kbt_p2}. A more detailed study, where $V/U$ is fixed and 
temperature is changed, shows that the SF phase at the central region does 
exist at lower temperatures. But, it melts to NF phase at the critical 
temperature of $k_{B}T/V = 0.16$. At the higher value of $V/U=0.5$ the central 
SF region re-emerges and so does the SS phase at still higher value of $V/U=1$. 
In short, with thermal fluctuations it is essential to have stronger dipolar
interactions to observe SS phase. Considering the parameters of the 
experimental realization of dipolar condensates of $^{168}$Er in optical 
lattices \cite{baier_16}, the corresponding temperature of 
$k_{B}T/V = 0.2$ is $\approx 40$ nK. This is within the experimental realm and
hence, the combined effect of dipolar interaction and artificial 
gauge field can lead to the emergence of SS phase within experimentally 
achievable parameter domain.

At higher temperatures, $k_{B}T/V \approx 0.3$, the central region is in NF
phase for weaker dipole interactions $V/U\leqslant 0.5$. Then, on increasing
$V/U$ further the density assumes checkerboard pattern, but the SF order
parameter is zero. That is the central region of the system is in the DW 
phase. This is to be compared and contrasted with the earlier result at 
$k_{B}T/V \approx 0.2$, where as shown in Fig.~\ref{dip_kbt_p2} the SS phase
exists for $V/U=1$. Thus, focusing on the strong interaction domain $V/U = 1$, 
our results show the existence of a SS-DW transition at $z~k_{B}T = V$. 
In short, the SS phase exists in the system at lower temperatures, but the
SS order melts into DW phase when $z~k_{B}T\geqslant V$. On increasing the
temperature further, the crystalline structure or diagonal long-range order 
of the DW phase starts to melt and at $z~k_{B}T\approx3~V$ system is fully in
the NF phase. So, the melting of the SS phase occurs in two steps. First, the 
off-diagonal long-range SF order is destroyed. This transforms the SS phase
into DW phase. And, second, the DW phase melts into NF phase. 
\begin{figure}[ht]
 {\includegraphics[width=9.0cm] {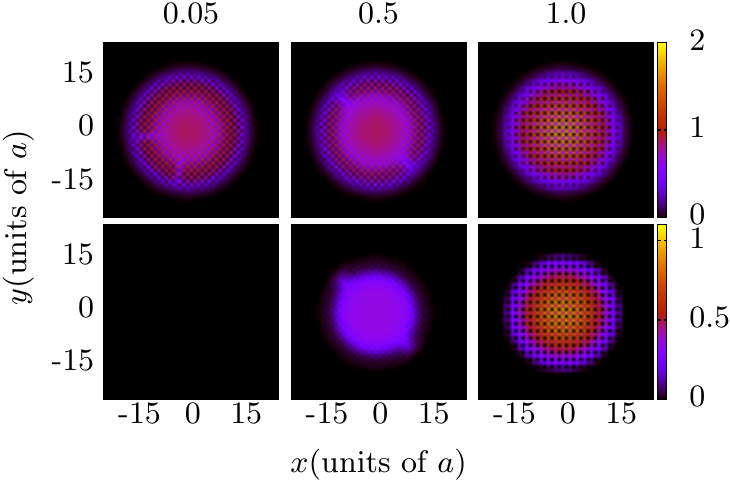}}
 \caption{The density distribution (upper panel) and SF order parameter
          (lower panel) of a $50\times 50$ square lattice in the presence
	  of artificial gauge field $\alpha = 1/2$ at $k_{B}T/V = 0.2$. 
	  The $V/U$ value is shown at the top of the figures. Here $x$ and 
	  $y$ are in units of the lattice spacing $a$.}
\label{dip_kbt_p2_alp5}
\end{figure}
In the presence of artificial gauge field the ground state and the corresponding
SF order parameter at $k_{B}T/V = 0.2$ are shown in Fig.~\ref{dip_kbt_p2_alp5}. 
As in the case of uniform system, there are no qualitative changes in the 
results with the introduction of artificial gauge field, although the atoms 
have velocity current. For the parameters considered, at $\mu/V = 2.8$ and 
$k_{B}T/V = 0.2$, the SS phase appears in the range $\Delta J=0.04$, whereas in
the presence of gauge field this range is enhanced to $\Delta J=0.07$. As 
discussed in the homogeneous case, for a given $\mu$ the range of $J$ for SS 
phase to occur is larger than without gauge field. This would enhance the 
possibility of observing the SS phase of eBHM in experimental realizations.


\section{Conclusions}
\label{conc}
We have examined the zero and finite temperature phase diagrams of eBHM in 
2D using SGMF and CGMF theory. In the presence of artificial gauge
field the domain of SS phase is enhanced and $2\times 2$ CGMF theory provide 
better description of the system. The cluster finite-size scaling 
analysis demonstrates the key role of fluctuations in determining the quantum 
phases. For the DW-SS transition, the suppression of fluctuations favours the 
DW phase. But, in the MI-SF transition, suppression of fluctuations favours 
the SF phase. This is indicated by the decrease and increase of $J_c$  
in the DW-SS and MI-SF transitions, respectively. At higher temperatures, the 
thermal fluctuations destroy the SS phase and the phase diagram exhibits 
NF-SF transition. Furthermore, we have studied the system of 
ultracold bosons with long range interactions in the presence of a harmonic 
confinement to relate with quantum gas experiments. Our results show that 
beyond a critical threshold of temperature $z~k_{B}T\geqslant V$, the SS 
phase vanishes and the system is occupied by the DW phase. These suggest that 
the prospect of observing the SS phase is higher when the temperature 
$z~k_{B}T < V$, this range of temperature is possible in the experiments of 
dipolar Bose gases loaded into optical lattices. This offers an opportunity 
to observe SS phase of eBHM in quantum dipolar gas experiments.   


\begin{acknowledgments}
The results presented in the paper are based on the computations using 
Vikram-100, the 100TFLOP HPC Cluster at Physical Research Laboratory, 
Ahmedabad, India. We thank Arko Roy, S. Gautam and S. A. Silotri for valuable 
discussions. K.S. acknowledges the support of the  National Science Centre, 
Poland via project 2016/21/B/ST2/01086.
\end{acknowledgments}


\bibliography{sup_sol}{}

\begin{thebibliography}{78}%
\makeatletter
\providecommand \@ifxundefined [1]{%
 \@ifx{#1\undefined}
}%
\providecommand \@ifnum [1]{%
 \ifnum #1\expandafter \@firstoftwo
 \else \expandafter \@secondoftwo
 \fi
}%
\providecommand \@ifx [1]{%
 \ifx #1\expandafter \@firstoftwo
 \else \expandafter \@secondoftwo
 \fi
}%
\providecommand \natexlab [1]{#1}%
\providecommand \enquote  [1]{``#1''}%
\providecommand \bibnamefont  [1]{#1}%
\providecommand \bibfnamefont [1]{#1}%
\providecommand \citenamefont [1]{#1}%
\providecommand \href@noop [0]{\@secondoftwo}%
\providecommand \href [0]{\begingroup \@sanitize@url \@href}%
\providecommand \@href[1]{\@@startlink{#1}\@@href}%
\providecommand \@@href[1]{\endgroup#1\@@endlink}%
\providecommand \@sanitize@url [0]{\catcode `\\12\catcode `\$12\catcode
  `\&12\catcode `\#12\catcode `\^12\catcode `\_12\catcode `\%12\relax}%
\providecommand \@@startlink[1]{}%
\providecommand \@@endlink[0]{}%
\providecommand \url  [0]{\begingroup\@sanitize@url \@url }%
\providecommand \@url [1]{\endgroup\@href {#1}{\urlprefix }}%
\providecommand \urlprefix  [0]{URL }%
\providecommand \Eprint [0]{\href }%
\providecommand \doibase [0]{http://dx.doi.org/}%
\providecommand \selectlanguage [0]{\@gobble}%
\providecommand \bibinfo  [0]{\@secondoftwo}%
\providecommand \bibfield  [0]{\@secondoftwo}%
\providecommand \translation [1]{[#1]}%
\providecommand \BibitemOpen [0]{}%
\providecommand \bibitemStop [0]{}%
\providecommand \bibitemNoStop [0]{.\EOS\space}%
\providecommand \EOS [0]{\spacefactor3000\relax}%
\providecommand \BibitemShut  [1]{\csname bibitem#1\endcsname}%
\let\auto@bib@innerbib\@empty
\bibitem [{\citenamefont {Lewenstein}\ \emph {et~al.}(2007)\citenamefont
  {Lewenstein}, \citenamefont {Sanpera}, \citenamefont {Ahufinger},
  \citenamefont {Damski}, \citenamefont {Sen(De)},\ and\ \citenamefont
  {Sen}}]{lewenstein_07}%
  \BibitemOpen
  \bibfield  {author} {\bibinfo {author} {\bibfnamefont {M.}~\bibnamefont
  {Lewenstein}}, \bibinfo {author} {\bibfnamefont {A.}~\bibnamefont {Sanpera}},
  \bibinfo {author} {\bibfnamefont {V.}~\bibnamefont {Ahufinger}}, \bibinfo
  {author} {\bibfnamefont {B.}~\bibnamefont {Damski}}, \bibinfo {author}
  {\bibfnamefont {A.}~\bibnamefont {Sen(De)}}, \ and\ \bibinfo {author}
  {\bibfnamefont {U.}~\bibnamefont {Sen}},\ }\bibfield  {title} {\enquote
  {\bibinfo {title} {Ultracold atomic gases in optical lattices: mimicking
  condensed matter physics and beyond},}\ }\href {\doibase
  10.1080/00018730701223200} {\bibfield  {journal} {\bibinfo  {journal} {Adv.
  Phys.}\ }\textbf {\bibinfo {volume} {56}},\ \bibinfo {pages} {243} (\bibinfo
  {year} {2007})}\BibitemShut {NoStop}%
\bibitem [{\citenamefont {Bloch}\ \emph {et~al.}(2008)\citenamefont {Bloch},
  \citenamefont {Dalibard},\ and\ \citenamefont {Zwerger}}]{bloch_08}%
  \BibitemOpen
  \bibfield  {author} {\bibinfo {author} {\bibfnamefont {I.}~\bibnamefont
  {Bloch}}, \bibinfo {author} {\bibfnamefont {J.}~\bibnamefont {Dalibard}}, \
  and\ \bibinfo {author} {\bibfnamefont {W.}~\bibnamefont {Zwerger}},\
  }\bibfield  {title} {\enquote {\bibinfo {title} {Many-body physics with
  ultracold gases},}\ }\href {\doibase 10.1103/RevModPhys.80.885} {\bibfield
  {journal} {\bibinfo  {journal} {Rev. Mod. Phys.}\ }\textbf {\bibinfo {volume}
  {80}},\ \bibinfo {pages} {885} (\bibinfo {year} {2008})}\BibitemShut
  {NoStop}%
\bibitem [{\citenamefont {Leggett}(1970)}]{leggett_70}%
  \BibitemOpen
  \bibfield  {author} {\bibinfo {author} {\bibfnamefont {A.~J.}\ \bibnamefont
  {Leggett}},\ }\bibfield  {title} {\enquote {\bibinfo {title} {Can a solid be
  superfluid?}}\ }\href {\doibase 10.1103/PhysRevLett.25.1543} {\bibfield
  {journal} {\bibinfo  {journal} {Phys. Rev. Lett.}\ }\textbf {\bibinfo
  {volume} {25}},\ \bibinfo {pages} {1543} (\bibinfo {year}
  {1970})}\BibitemShut {NoStop}%
\bibitem [{\citenamefont {Chester}(1970)}]{chester_70}%
  \BibitemOpen
  \bibfield  {author} {\bibinfo {author} {\bibfnamefont {G.~V.}\ \bibnamefont
  {Chester}},\ }\bibfield  {title} {\enquote {\bibinfo {title} {Speculations on
  {B}ose-{E}instein condensation and quantum crystals},}\ }\href {\doibase
  10.1103/PhysRevA.2.256} {\bibfield  {journal} {\bibinfo  {journal} {Phys.
  Rev. A}\ }\textbf {\bibinfo {volume} {2}},\ \bibinfo {pages} {256} (\bibinfo
  {year} {1970})}\BibitemShut {NoStop}%
\bibitem [{\citenamefont {Andreev}\ and\ \citenamefont
  {Lifshitz}(1969)}]{andreev_69}%
  \BibitemOpen
  \bibfield  {author} {\bibinfo {author} {\bibfnamefont {A.~F.}\ \bibnamefont
  {Andreev}}\ and\ \bibinfo {author} {\bibfnamefont {I.~M.}\ \bibnamefont
  {Lifshitz}},\ }\bibfield  {title} {\enquote {\bibinfo {title} {Quantum theory
  of defects in crystals},}\ }\href@noop {} {\bibfield  {journal} {\bibinfo
  {journal} {Sov. Phys. JETP}\ }\textbf {\bibinfo {volume} {29}},\ \bibinfo
  {pages} {1107} (\bibinfo {year} {1969})}\BibitemShut {NoStop}%
\bibitem [{\citenamefont {Thouless}(1969)}]{thouless_69}%
  \BibitemOpen
  \bibfield  {author} {\bibinfo {author} {\bibfnamefont {D.~J.}\ \bibnamefont
  {Thouless}},\ }\bibfield  {title} {\enquote {\bibinfo {title} {The flow of a
  dense superfluid},}\ }\href {\doibase 10.1016/0003-4916(69)90286-3}
  {\bibfield  {journal} {\bibinfo  {journal} {Ann. Phys.}\ }\textbf {\bibinfo
  {volume} {52}},\ \bibinfo {pages} {403} (\bibinfo {year} {1969})}\BibitemShut
  {NoStop}%
\bibitem [{\citenamefont {Kim}\ and\ \citenamefont
  {Chan}(2004{\natexlab{a}})}]{kim_04}%
  \BibitemOpen
  \bibfield  {author} {\bibinfo {author} {\bibfnamefont {E.}~\bibnamefont
  {Kim}}\ and\ \bibinfo {author} {\bibfnamefont {M.~H.~W.}\ \bibnamefont
  {Chan}},\ }\bibfield  {title} {\enquote {\bibinfo {title} {Probable
  observation of a supersolid helium phase},}\ }\href {\doibase
  10.1038/nature02220} {\bibfield  {journal} {\bibinfo  {journal} {Nature
  (London)}\ }\textbf {\bibinfo {volume} {427}},\ \bibinfo {pages} {225}
  (\bibinfo {year} {2004}{\natexlab{a}})}\BibitemShut {NoStop}%
\bibitem [{\citenamefont {Kim}\ and\ \citenamefont
  {Chan}(2004{\natexlab{b}})}]{kim_04a}%
  \BibitemOpen
  \bibfield  {author} {\bibinfo {author} {\bibfnamefont {E.}~\bibnamefont
  {Kim}}\ and\ \bibinfo {author} {\bibfnamefont {M.~H.~W.}\ \bibnamefont
  {Chan}},\ }\bibfield  {title} {\enquote {\bibinfo {title} {Observation of
  superflow in solid helium},}\ }\href {\doibase 10.1126/science.1101501}
  {\bibfield  {journal} {\bibinfo  {journal} {Science}\ }\textbf {\bibinfo
  {volume} {305}},\ \bibinfo {pages} {1941} (\bibinfo {year}
  {2004}{\natexlab{b}})}\BibitemShut {NoStop}%
\bibitem [{\citenamefont {Kim}\ and\ \citenamefont {Chan}(2012)}]{kim_12}%
  \BibitemOpen
  \bibfield  {author} {\bibinfo {author} {\bibfnamefont {D.~Y.}\ \bibnamefont
  {Kim}}\ and\ \bibinfo {author} {\bibfnamefont {M.~H.~W.}\ \bibnamefont
  {Chan}},\ }\bibfield  {title} {\enquote {\bibinfo {title} {Absence of
  supersolidity in solid helium in porous vycor glass},}\ }\href {\doibase
  10.1103/PhysRevLett.109.155301} {\bibfield  {journal} {\bibinfo  {journal}
  {Phys. Rev. Lett.}\ }\textbf {\bibinfo {volume} {109}},\ \bibinfo {pages}
  {155301} (\bibinfo {year} {2012})}\BibitemShut {NoStop}%
\bibitem [{\citenamefont {Léonard}\ \emph {et~al.}(2017)\citenamefont
  {Léonard}, \citenamefont {Morales}, \citenamefont {Zupancic}, \citenamefont
  {Esslinger},\ and\ \citenamefont {Donner}}]{leonard_17}%
  \BibitemOpen
  \bibfield  {author} {\bibinfo {author} {\bibfnamefont {J.}~\bibnamefont
  {Léonard}}, \bibinfo {author} {\bibfnamefont {A.}~\bibnamefont {Morales}},
  \bibinfo {author} {\bibfnamefont {P.}~\bibnamefont {Zupancic}}, \bibinfo
  {author} {\bibfnamefont {T.}~\bibnamefont {Esslinger}}, \ and\ \bibinfo
  {author} {\bibfnamefont {T.}~\bibnamefont {Donner}},\ }\bibfield  {title}
  {\enquote {\bibinfo {title} {Supersolid formation in a quantum gas breaking a
  continuous translational symmetry},}\ }\href {\doibase 10.1038/nature21067}
  {\bibfield  {journal} {\bibinfo  {journal} {Nature (London)}\ }\textbf
  {\bibinfo {volume} {543}},\ \bibinfo {pages} {87} (\bibinfo {year}
  {2017})}\BibitemShut {NoStop}%
\bibitem [{\citenamefont {Li}\ \emph {et~al.}(2017)\citenamefont {Li},
  \citenamefont {Lee}, \citenamefont {Huang}, \citenamefont {Burchesky},
  \citenamefont {Shteynas}, \citenamefont {Top}, \citenamefont {Jamison},\ and\
  \citenamefont {Ketterle}}]{li_17}%
  \BibitemOpen
  \bibfield  {author} {\bibinfo {author} {\bibfnamefont {J.-R.}\ \bibnamefont
  {Li}}, \bibinfo {author} {\bibfnamefont {J.}~\bibnamefont {Lee}}, \bibinfo
  {author} {\bibfnamefont {W.}~\bibnamefont {Huang}}, \bibinfo {author}
  {\bibfnamefont {S.}~\bibnamefont {Burchesky}}, \bibinfo {author}
  {\bibfnamefont {B.}~\bibnamefont {Shteynas}}, \bibinfo {author}
  {\bibfnamefont {F.~Ç.}\ \bibnamefont {Top}}, \bibinfo {author}
  {\bibfnamefont {A.~O.}\ \bibnamefont {Jamison}}, \ and\ \bibinfo {author}
  {\bibfnamefont {W.}~\bibnamefont {Ketterle}},\ }\bibfield  {title} {\enquote
  {\bibinfo {title} {A stripe phase with supersolid properties in
  spin-orbit-coupled {B}ose-{E}instein condensates},}\ }\href {\doibase
  10.1038/nature21431} {\bibfield  {journal} {\bibinfo  {journal} {Nature
  (London)}\ }\textbf {\bibinfo {volume} {543}},\ \bibinfo {pages} {91}
  (\bibinfo {year} {2017})}\BibitemShut {NoStop}%
\bibitem [{\citenamefont {Tanzi}\ \emph
  {et~al.}(2019{\natexlab{a}})\citenamefont {Tanzi}, \citenamefont {Lucioni},
  \citenamefont {Fam\`a}, \citenamefont {Catani}, \citenamefont {Fioretti},
  \citenamefont {Gabbanini}, \citenamefont {Bisset}, \citenamefont {Santos},\
  and\ \citenamefont {Modugno}}]{tanzi_19}%
  \BibitemOpen
  \bibfield  {author} {\bibinfo {author} {\bibfnamefont {L.}~\bibnamefont
  {Tanzi}}, \bibinfo {author} {\bibfnamefont {E.}~\bibnamefont {Lucioni}},
  \bibinfo {author} {\bibfnamefont {F.}~\bibnamefont {Fam\`a}}, \bibinfo
  {author} {\bibfnamefont {J.}~\bibnamefont {Catani}}, \bibinfo {author}
  {\bibfnamefont {A.}~\bibnamefont {Fioretti}}, \bibinfo {author}
  {\bibfnamefont {C.}~\bibnamefont {Gabbanini}}, \bibinfo {author}
  {\bibfnamefont {R.~N.}\ \bibnamefont {Bisset}}, \bibinfo {author}
  {\bibfnamefont {L.}~\bibnamefont {Santos}}, \ and\ \bibinfo {author}
  {\bibfnamefont {G.}~\bibnamefont {Modugno}},\ }\bibfield  {title} {\enquote
  {\bibinfo {title} {Observation of a dipolar quantum gas with metastable
  supersolid properties},}\ }\href {\doibase 10.1103/PhysRevLett.122.130405}
  {\bibfield  {journal} {\bibinfo  {journal} {Phys. Rev. Lett.}\ }\textbf
  {\bibinfo {volume} {122}},\ \bibinfo {pages} {130405} (\bibinfo {year}
  {2019}{\natexlab{a}})}\BibitemShut {NoStop}%
\bibitem [{\citenamefont {B\"ottcher}\ \emph {et~al.}(2019)\citenamefont
  {B\"ottcher}, \citenamefont {Schmidt}, \citenamefont {Wenzel}, \citenamefont
  {Hertkorn}, \citenamefont {Guo}, \citenamefont {Langen},\ and\ \citenamefont
  {Pfau}}]{bottcher_19}%
  \BibitemOpen
  \bibfield  {author} {\bibinfo {author} {\bibfnamefont {F.}~\bibnamefont
  {B\"ottcher}}, \bibinfo {author} {\bibfnamefont {J.-N.}\ \bibnamefont
  {Schmidt}}, \bibinfo {author} {\bibfnamefont {M.}~\bibnamefont {Wenzel}},
  \bibinfo {author} {\bibfnamefont {J.}~\bibnamefont {Hertkorn}}, \bibinfo
  {author} {\bibfnamefont {M.}~\bibnamefont {Guo}}, \bibinfo {author}
  {\bibfnamefont {T.}~\bibnamefont {Langen}}, \ and\ \bibinfo {author}
  {\bibfnamefont {T.}~\bibnamefont {Pfau}},\ }\bibfield  {title} {\enquote
  {\bibinfo {title} {Transient supersolid properties in an array of dipolar
  quantum droplets},}\ }\href {\doibase 10.1103/PhysRevX.9.011051} {\bibfield
  {journal} {\bibinfo  {journal} {Phys. Rev. X}\ }\textbf {\bibinfo {volume}
  {9}},\ \bibinfo {pages} {011051} (\bibinfo {year} {2019})}\BibitemShut
  {NoStop}%
\bibitem [{\citenamefont {Chomaz}\ \emph {et~al.}(2019)\citenamefont {Chomaz},
  \citenamefont {Petter}, \citenamefont {Ilzh\"ofer}, \citenamefont {Natale},
  \citenamefont {Trautmann}, \citenamefont {Politi}, \citenamefont
  {Durastante}, \citenamefont {van Bijnen}, \citenamefont {Patscheider},
  \citenamefont {Sohmen}, \citenamefont {Mark},\ and\ \citenamefont
  {Ferlaino}}]{chomaz_19}%
  \BibitemOpen
  \bibfield  {author} {\bibinfo {author} {\bibfnamefont {L.}~\bibnamefont
  {Chomaz}}, \bibinfo {author} {\bibfnamefont {D.}~\bibnamefont {Petter}},
  \bibinfo {author} {\bibfnamefont {P.}~\bibnamefont {Ilzh\"ofer}}, \bibinfo
  {author} {\bibfnamefont {G.}~\bibnamefont {Natale}}, \bibinfo {author}
  {\bibfnamefont {A.}~\bibnamefont {Trautmann}}, \bibinfo {author}
  {\bibfnamefont {C.}~\bibnamefont {Politi}}, \bibinfo {author} {\bibfnamefont
  {G.}~\bibnamefont {Durastante}}, \bibinfo {author} {\bibfnamefont {R.~M.~W.}\
  \bibnamefont {van Bijnen}}, \bibinfo {author} {\bibfnamefont
  {A.}~\bibnamefont {Patscheider}}, \bibinfo {author} {\bibfnamefont
  {M.}~\bibnamefont {Sohmen}}, \bibinfo {author} {\bibfnamefont {M.~J.}\
  \bibnamefont {Mark}}, \ and\ \bibinfo {author} {\bibfnamefont
  {F.}~\bibnamefont {Ferlaino}},\ }\bibfield  {title} {\enquote {\bibinfo
  {title} {Long-lived and transient supersolid behaviors in dipolar quantum
  gases},}\ }\href {\doibase 10.1103/PhysRevX.9.021012} {\bibfield  {journal}
  {\bibinfo  {journal} {Phys. Rev. X}\ }\textbf {\bibinfo {volume} {9}},\
  \bibinfo {pages} {021012} (\bibinfo {year} {2019})}\BibitemShut {NoStop}%
\bibitem [{\citenamefont {Landig}\ \emph {et~al.}(2016)\citenamefont {Landig},
  \citenamefont {Hruby}, \citenamefont {Dogra}, \citenamefont {Landini},
  \citenamefont {Mottl}, \citenamefont {Donner},\ and\ \citenamefont
  {Esslinger}}]{landig_16}%
  \BibitemOpen
  \bibfield  {author} {\bibinfo {author} {\bibfnamefont {R.}~\bibnamefont
  {Landig}}, \bibinfo {author} {\bibfnamefont {L.}~\bibnamefont {Hruby}},
  \bibinfo {author} {\bibfnamefont {N.}~\bibnamefont {Dogra}}, \bibinfo
  {author} {\bibfnamefont {M.}~\bibnamefont {Landini}}, \bibinfo {author}
  {\bibfnamefont {R.}~\bibnamefont {Mottl}}, \bibinfo {author} {\bibfnamefont
  {T.}~\bibnamefont {Donner}}, \ and\ \bibinfo {author} {\bibfnamefont
  {T.}~\bibnamefont {Esslinger}},\ }\bibfield  {title} {\enquote {\bibinfo
  {title} {Quantum phases from competing short- and long-range interactions in
  an optical lattice},}\ }\href {\doibase 10.1038/nature17409} {\bibfield
  {journal} {\bibinfo  {journal} {Nature (London)}\ }\textbf {\bibinfo {volume}
  {532}},\ \bibinfo {pages} {476} (\bibinfo {year} {2016})}\BibitemShut
  {NoStop}%
\bibitem [{\citenamefont {Natale}\ \emph {et~al.}(2019)\citenamefont {Natale},
  \citenamefont {van Bijnen}, \citenamefont {Patscheider}, \citenamefont
  {Petter}, \citenamefont {Mark}, \citenamefont {Chomaz},\ and\ \citenamefont
  {Ferlaino}}]{natale_19}%
  \BibitemOpen
  \bibfield  {author} {\bibinfo {author} {\bibfnamefont {G.}~\bibnamefont
  {Natale}}, \bibinfo {author} {\bibfnamefont {R.~M.~W.}\ \bibnamefont {van
  Bijnen}}, \bibinfo {author} {\bibfnamefont {A.}~\bibnamefont {Patscheider}},
  \bibinfo {author} {\bibfnamefont {D.}~\bibnamefont {Petter}}, \bibinfo
  {author} {\bibfnamefont {M.~J.}\ \bibnamefont {Mark}}, \bibinfo {author}
  {\bibfnamefont {L.}~\bibnamefont {Chomaz}}, \ and\ \bibinfo {author}
  {\bibfnamefont {F.}~\bibnamefont {Ferlaino}},\ }\bibfield  {title} {\enquote
  {\bibinfo {title} {Excitation spectrum of a trapped dipolar supersolid and
  its experimental evidence},}\ }\href {\doibase
  10.1103/PhysRevLett.123.050402} {\bibfield  {journal} {\bibinfo  {journal}
  {Phys. Rev. Lett.}\ }\textbf {\bibinfo {volume} {123}},\ \bibinfo {pages}
  {050402} (\bibinfo {year} {2019})}\BibitemShut {NoStop}%
\bibitem [{\citenamefont {Tanzi}\ \emph
  {et~al.}(2019{\natexlab{b}})\citenamefont {Tanzi}, \citenamefont {Roccuzzo},
  \citenamefont {Lucioni}, \citenamefont {Famà}, \citenamefont {Fioretti},
  \citenamefont {Gabbanini}, \citenamefont {Modugno}, \citenamefont {Recati},\
  and\ \citenamefont {Stringari}}]{tanzi_19a}%
  \BibitemOpen
  \bibfield  {author} {\bibinfo {author} {\bibfnamefont {L.}~\bibnamefont
  {Tanzi}}, \bibinfo {author} {\bibfnamefont {S.~M.}\ \bibnamefont {Roccuzzo}},
  \bibinfo {author} {\bibfnamefont {E.}~\bibnamefont {Lucioni}}, \bibinfo
  {author} {\bibfnamefont {F.}~\bibnamefont {Famà}}, \bibinfo {author}
  {\bibfnamefont {A.}~\bibnamefont {Fioretti}}, \bibinfo {author}
  {\bibfnamefont {C.}~\bibnamefont {Gabbanini}}, \bibinfo {author}
  {\bibfnamefont {G.}~\bibnamefont {Modugno}}, \bibinfo {author} {\bibfnamefont
  {A.}~\bibnamefont {Recati}}, \ and\ \bibinfo {author} {\bibfnamefont
  {S.}~\bibnamefont {Stringari}},\ }\bibfield  {title} {\enquote {\bibinfo
  {title} {Supersolid symmetry breaking from compressional oscillations in a
  dipolar quantum gas},}\ }\href {\doibase 10.1038/s41586-019-1568-6}
  {\bibfield  {journal} {\bibinfo  {journal} {Nature}\ }\textbf {\bibinfo
  {volume} {574}},\ \bibinfo {pages} {382} (\bibinfo {year}
  {2019}{\natexlab{b}})}\BibitemShut {NoStop}%
\bibitem [{\citenamefont {Guo}\ \emph {et~al.}(2019)\citenamefont {Guo},
  \citenamefont {Böttcher}, \citenamefont {Hertkorn}, \citenamefont {Schmidt},
  \citenamefont {Wenzel}, \citenamefont {Büchler}, \citenamefont {Langen},\
  and\ \citenamefont {Pfau}}]{guo_19}%
  \BibitemOpen
  \bibfield  {author} {\bibinfo {author} {\bibfnamefont {M.}~\bibnamefont
  {Guo}}, \bibinfo {author} {\bibfnamefont {F.}~\bibnamefont {Böttcher}},
  \bibinfo {author} {\bibfnamefont {J.}~\bibnamefont {Hertkorn}}, \bibinfo
  {author} {\bibfnamefont {J.-N.}\ \bibnamefont {Schmidt}}, \bibinfo {author}
  {\bibfnamefont {M.}~\bibnamefont {Wenzel}}, \bibinfo {author} {\bibfnamefont
  {H.~P.}\ \bibnamefont {Büchler}}, \bibinfo {author} {\bibfnamefont
  {T.}~\bibnamefont {Langen}}, \ and\ \bibinfo {author} {\bibfnamefont
  {T.}~\bibnamefont {Pfau}},\ }\bibfield  {title} {\enquote {\bibinfo {title}
  {The low-energy goldstone mode in a trapped dipolar supersolid},}\ }\href
  {\doibase 10.1038/s41586-019-1569-5} {\bibfield  {journal} {\bibinfo
  {journal} {Nature}\ }\textbf {\bibinfo {volume} {574}},\ \bibinfo {pages}
  {386} (\bibinfo {year} {2019})}\BibitemShut {NoStop}%
\bibitem [{\citenamefont {Batrouni}\ and\ \citenamefont
  {Scalettar}(2000)}]{batrouni_00}%
  \BibitemOpen
  \bibfield  {author} {\bibinfo {author} {\bibfnamefont {G.~G.}\ \bibnamefont
  {Batrouni}}\ and\ \bibinfo {author} {\bibfnamefont {R.~T.}\ \bibnamefont
  {Scalettar}},\ }\bibfield  {title} {\enquote {\bibinfo {title} {Phase
  separation in supersolids},}\ }\href {\doibase 10.1103/PhysRevLett.84.1599}
  {\bibfield  {journal} {\bibinfo  {journal} {Phys. Rev. Lett.}\ }\textbf
  {\bibinfo {volume} {84}},\ \bibinfo {pages} {1599} (\bibinfo {year}
  {2000})}\BibitemShut {NoStop}%
\bibitem [{\citenamefont {van Otterlo}\ \emph {et~al.}(1995)\citenamefont {van
  Otterlo}, \citenamefont {Wagenblast}, \citenamefont {Baltin}, \citenamefont
  {Bruder}, \citenamefont {Fazio},\ and\ \citenamefont {Sch\"on}}]{otterlo_95}%
  \BibitemOpen
  \bibfield  {author} {\bibinfo {author} {\bibfnamefont {Anne}\ \bibnamefont
  {van Otterlo}}, \bibinfo {author} {\bibfnamefont {Karl-Heinz}\ \bibnamefont
  {Wagenblast}}, \bibinfo {author} {\bibfnamefont {Reinhard}\ \bibnamefont
  {Baltin}}, \bibinfo {author} {\bibfnamefont {C.}~\bibnamefont {Bruder}},
  \bibinfo {author} {\bibfnamefont {Rosario}\ \bibnamefont {Fazio}}, \ and\
  \bibinfo {author} {\bibfnamefont {Gerd}\ \bibnamefont {Sch\"on}},\ }\bibfield
   {title} {\enquote {\bibinfo {title} {Quantum phase transitions of
  interacting bosons and the supersolid phase},}\ }\href {\doibase
  10.1103/PhysRevB.52.16176} {\bibfield  {journal} {\bibinfo  {journal} {Phys.
  Rev. B}\ }\textbf {\bibinfo {volume} {52}},\ \bibinfo {pages} {16176--16186}
  (\bibinfo {year} {1995})}\BibitemShut {NoStop}%
\bibitem [{\citenamefont {Sengupta}\ \emph {et~al.}(2005)\citenamefont
  {Sengupta}, \citenamefont {Pryadko}, \citenamefont {Alet}, \citenamefont
  {Troyer},\ and\ \citenamefont {Schmid}}]{sengupta_05}%
  \BibitemOpen
  \bibfield  {author} {\bibinfo {author} {\bibfnamefont {P.}~\bibnamefont
  {Sengupta}}, \bibinfo {author} {\bibfnamefont {L.~P.}\ \bibnamefont
  {Pryadko}}, \bibinfo {author} {\bibfnamefont {F.}~\bibnamefont {Alet}},
  \bibinfo {author} {\bibfnamefont {M.}~\bibnamefont {Troyer}}, \ and\ \bibinfo
  {author} {\bibfnamefont {G.}~\bibnamefont {Schmid}},\ }\bibfield  {title}
  {\enquote {\bibinfo {title} {Supersolids versus phase separation in
  two-dimensional lattice bosons},}\ }\href {\doibase
  10.1103/PhysRevLett.94.207202} {\bibfield  {journal} {\bibinfo  {journal}
  {Phys. Rev. Lett.}\ }\textbf {\bibinfo {volume} {94}},\ \bibinfo {pages}
  {207202} (\bibinfo {year} {2005})}\BibitemShut {NoStop}%
\bibitem [{\citenamefont {Ohgoe}\ \emph
  {et~al.}(2012{\natexlab{a}})\citenamefont {Ohgoe}, \citenamefont {Suzuki},\
  and\ \citenamefont {Kawashima}}]{ohgoe_12b}%
  \BibitemOpen
  \bibfield  {author} {\bibinfo {author} {\bibfnamefont {Takahiro}\
  \bibnamefont {Ohgoe}}, \bibinfo {author} {\bibfnamefont {Takafumi}\
  \bibnamefont {Suzuki}}, \ and\ \bibinfo {author} {\bibfnamefont {Naoki}\
  \bibnamefont {Kawashima}},\ }\bibfield  {title} {\enquote {\bibinfo {title}
  {Ground-state phase diagram of the two-dimensional extended bose-hubbard
  model},}\ }\href {\doibase 10.1103/PhysRevB.86.054520} {\bibfield  {journal}
  {\bibinfo  {journal} {Phys. Rev. B}\ }\textbf {\bibinfo {volume} {86}},\
  \bibinfo {pages} {054520} (\bibinfo {year} {2012}{\natexlab{a}})}\BibitemShut
  {NoStop}%
\bibitem [{\citenamefont {Wessel}(2007)}]{wessel_07}%
  \BibitemOpen
  \bibfield  {author} {\bibinfo {author} {\bibfnamefont {S.}~\bibnamefont
  {Wessel}},\ }\bibfield  {title} {\enquote {\bibinfo {title} {Phase diagram of
  interacting bosons on the honeycomb lattice},}\ }\href {\doibase
  10.1103/PhysRevB.75.174301} {\bibfield  {journal} {\bibinfo  {journal} {Phys.
  Rev. B}\ }\textbf {\bibinfo {volume} {75}},\ \bibinfo {pages} {174301}
  (\bibinfo {year} {2007})}\BibitemShut {NoStop}%
\bibitem [{\citenamefont {Gan}\ \emph {et~al.}(2007)\citenamefont {Gan},
  \citenamefont {Wen}, \citenamefont {Ye}, \citenamefont {Li}, \citenamefont
  {Yang},\ and\ \citenamefont {Yu}}]{gan_07}%
  \BibitemOpen
  \bibfield  {author} {\bibinfo {author} {\bibfnamefont {J.~Y.}\ \bibnamefont
  {Gan}}, \bibinfo {author} {\bibfnamefont {Y.~C.}\ \bibnamefont {Wen}},
  \bibinfo {author} {\bibfnamefont {J.}~\bibnamefont {Ye}}, \bibinfo {author}
  {\bibfnamefont {T.}~\bibnamefont {Li}}, \bibinfo {author} {\bibfnamefont
  {S.-J.}\ \bibnamefont {Yang}}, \ and\ \bibinfo {author} {\bibfnamefont
  {Y.}~\bibnamefont {Yu}},\ }\bibfield  {title} {\enquote {\bibinfo {title}
  {Extended {B}ose-{H}ubbard model on a honeycomb lattice},}\ }\href {\doibase
  10.1103/PhysRevB.75.214509} {\bibfield  {journal} {\bibinfo  {journal} {Phys.
  Rev. B}\ }\textbf {\bibinfo {volume} {75}},\ \bibinfo {pages} {214509}
  (\bibinfo {year} {2007})}\BibitemShut {NoStop}%
\bibitem [{\citenamefont {Flottat}\ \emph {et~al.}(2017)\citenamefont
  {Flottat}, \citenamefont {deParny}, \citenamefont {H\'ebert}, \citenamefont
  {Rousseau},\ and\ \citenamefont {Batrouni}}]{flottat_17}%
  \BibitemOpen
  \bibfield  {author} {\bibinfo {author} {\bibfnamefont {T.}~\bibnamefont
  {Flottat}}, \bibinfo {author} {\bibfnamefont {L.~deForges}\ \bibnamefont
  {deParny}}, \bibinfo {author} {\bibfnamefont {F.}~\bibnamefont {H\'ebert}},
  \bibinfo {author} {\bibfnamefont {V.~G.}\ \bibnamefont {Rousseau}}, \ and\
  \bibinfo {author} {\bibfnamefont {G.~G.}\ \bibnamefont {Batrouni}},\
  }\bibfield  {title} {\enquote {\bibinfo {title} {Phase diagram of bosons in a
  two-dimensional optical lattice with infinite-range cavity-mediated
  interactions},}\ }\href {\doibase 10.1103/PhysRevB.95.144501} {\bibfield
  {journal} {\bibinfo  {journal} {Phys. Rev. B}\ }\textbf {\bibinfo {volume}
  {95}},\ \bibinfo {pages} {144501} (\bibinfo {year} {2017})}\BibitemShut
  {NoStop}%
\bibitem [{\citenamefont {Mathey}\ \emph {et~al.}(2009)\citenamefont {Mathey},
  \citenamefont {Danshita},\ and\ \citenamefont {Clark}}]{mathey_09}%
  \BibitemOpen
  \bibfield  {author} {\bibinfo {author} {\bibfnamefont {L.}~\bibnamefont
  {Mathey}}, \bibinfo {author} {\bibfnamefont {I.}~\bibnamefont {Danshita}}, \
  and\ \bibinfo {author} {\bibfnamefont {C.~W.}\ \bibnamefont {Clark}},\
  }\bibfield  {title} {\enquote {\bibinfo {title} {Creating a supersolid in
  one-dimensional {B}ose mixtures},}\ }\href {\doibase
  10.1103/PhysRevA.79.011602} {\bibfield  {journal} {\bibinfo  {journal} {Phys.
  Rev. A}\ }\textbf {\bibinfo {volume} {79}},\ \bibinfo {pages} {011602(R)}
  (\bibinfo {year} {2009})}\BibitemShut {NoStop}%
\bibitem [{\citenamefont {Batrouni}\ \emph {et~al.}(2006)\citenamefont
  {Batrouni}, \citenamefont {H\'ebert},\ and\ \citenamefont
  {Scalettar}}]{batrouni_06}%
  \BibitemOpen
  \bibfield  {author} {\bibinfo {author} {\bibfnamefont {G.~G.}\ \bibnamefont
  {Batrouni}}, \bibinfo {author} {\bibfnamefont {F.}~\bibnamefont {H\'ebert}},
  \ and\ \bibinfo {author} {\bibfnamefont {R.~T.}\ \bibnamefont {Scalettar}},\
  }\bibfield  {title} {\enquote {\bibinfo {title} {Supersolid phases in the
  one-dimensional extended soft-core bosonic {H}ubbard model},}\ }\href
  {\doibase 10.1103/PhysRevLett.97.087209} {\bibfield  {journal} {\bibinfo
  {journal} {Phys. Rev. Lett.}\ }\textbf {\bibinfo {volume} {97}},\ \bibinfo
  {pages} {087209} (\bibinfo {year} {2006})}\BibitemShut {NoStop}%
\bibitem [{\citenamefont {Sachdeva}\ \emph {et~al.}(2017)\citenamefont
  {Sachdeva}, \citenamefont {Singh},\ and\ \citenamefont
  {Busch}}]{sachdeva_17}%
  \BibitemOpen
  \bibfield  {author} {\bibinfo {author} {\bibfnamefont {R.}~\bibnamefont
  {Sachdeva}}, \bibinfo {author} {\bibfnamefont {M.}~\bibnamefont {Singh}}, \
  and\ \bibinfo {author} {\bibfnamefont {T.}~\bibnamefont {Busch}},\ }\bibfield
   {title} {\enquote {\bibinfo {title} {Extended {B}ose-{H}ubbard model for
  two-leg ladder systems in artificial magnetic fields},}\ }\href {\doibase
  10.1103/PhysRevA.95.063601} {\bibfield  {journal} {\bibinfo  {journal} {Phys.
  Rev. A}\ }\textbf {\bibinfo {volume} {95}},\ \bibinfo {pages} {063601}
  (\bibinfo {year} {2017})}\BibitemShut {NoStop}%
\bibitem [{\citenamefont {G\'oral}\ \emph {et~al.}(2002)\citenamefont
  {G\'oral}, \citenamefont {Santos},\ and\ \citenamefont
  {Lewenstein}}]{goral_02}%
  \BibitemOpen
  \bibfield  {author} {\bibinfo {author} {\bibfnamefont {K.}~\bibnamefont
  {G\'oral}}, \bibinfo {author} {\bibfnamefont {L.}~\bibnamefont {Santos}}, \
  and\ \bibinfo {author} {\bibfnamefont {M.}~\bibnamefont {Lewenstein}},\
  }\bibfield  {title} {\enquote {\bibinfo {title} {Quantum phases of dipolar
  bosons in optical lattices},}\ }\href {\doibase
  10.1103/PhysRevLett.88.170406} {\bibfield  {journal} {\bibinfo  {journal}
  {Phys. Rev. Lett.}\ }\textbf {\bibinfo {volume} {88}},\ \bibinfo {pages}
  {170406} (\bibinfo {year} {2002})}\BibitemShut {NoStop}%
\bibitem [{\citenamefont {Scarola}\ and\ \citenamefont
  {Das~Sarma}(2005)}]{scarola_05}%
  \BibitemOpen
  \bibfield  {author} {\bibinfo {author} {\bibfnamefont {V.~W.}\ \bibnamefont
  {Scarola}}\ and\ \bibinfo {author} {\bibfnamefont {S.}~\bibnamefont
  {Das~Sarma}},\ }\bibfield  {title} {\enquote {\bibinfo {title} {Quantum
  phases of the extended {B}ose-{H}ubbard {H}amiltonian: {P}ossibility of a
  supersolid state of cold atoms in optical lattices},}\ }\href {\doibase
  10.1103/PhysRevLett.95.033003} {\bibfield  {journal} {\bibinfo  {journal}
  {Phys. Rev. Lett.}\ }\textbf {\bibinfo {volume} {95}},\ \bibinfo {pages}
  {033003} (\bibinfo {year} {2005})}\BibitemShut {NoStop}%
\bibitem [{\citenamefont {Yi}\ \emph {et~al.}(2007)\citenamefont {Yi},
  \citenamefont {Li},\ and\ \citenamefont {Sun}}]{yi_07}%
  \BibitemOpen
  \bibfield  {author} {\bibinfo {author} {\bibfnamefont {S.}~\bibnamefont
  {Yi}}, \bibinfo {author} {\bibfnamefont {T.}~\bibnamefont {Li}}, \ and\
  \bibinfo {author} {\bibfnamefont {C.~P.}\ \bibnamefont {Sun}},\ }\bibfield
  {title} {\enquote {\bibinfo {title} {Novel quantum phases of dipolar {B}ose
  gases in optical lattices},}\ }\href {\doibase 10.1103/PhysRevLett.98.260405}
  {\bibfield  {journal} {\bibinfo  {journal} {Phys. Rev. Lett.}\ }\textbf
  {\bibinfo {volume} {98}},\ \bibinfo {pages} {260405} (\bibinfo {year}
  {2007})}\BibitemShut {NoStop}%
\bibitem [{\citenamefont {Ng}\ and\ \citenamefont {Chen}(2008)}]{ng_08}%
  \BibitemOpen
  \bibfield  {author} {\bibinfo {author} {\bibfnamefont {K.-K.}\ \bibnamefont
  {Ng}}\ and\ \bibinfo {author} {\bibfnamefont {Y.-C.}\ \bibnamefont {Chen}},\
  }\bibfield  {title} {\enquote {\bibinfo {title} {Supersolid phases in the
  bosonic extended {H}ubbard model},}\ }\href {\doibase
  10.1103/PhysRevB.77.052506} {\bibfield  {journal} {\bibinfo  {journal} {Phys.
  Rev. B}\ }\textbf {\bibinfo {volume} {77}},\ \bibinfo {pages} {052506}
  (\bibinfo {year} {2008})}\BibitemShut {NoStop}%
\bibitem [{\citenamefont {Capogrosso-Sansone}\ \emph
  {et~al.}(2010)\citenamefont {Capogrosso-Sansone}, \citenamefont {Trefzger},
  \citenamefont {Lewenstein}, \citenamefont {Zoller},\ and\ \citenamefont
  {Pupillo}}]{capogrosso_10}%
  \BibitemOpen
  \bibfield  {author} {\bibinfo {author} {\bibfnamefont {B.}~\bibnamefont
  {Capogrosso-Sansone}}, \bibinfo {author} {\bibfnamefont {C.}~\bibnamefont
  {Trefzger}}, \bibinfo {author} {\bibfnamefont {M.}~\bibnamefont
  {Lewenstein}}, \bibinfo {author} {\bibfnamefont {P.}~\bibnamefont {Zoller}},
  \ and\ \bibinfo {author} {\bibfnamefont {G.}~\bibnamefont {Pupillo}},\
  }\bibfield  {title} {\enquote {\bibinfo {title} {Quantum phases of cold polar
  molecules in {2D} optical lattices},}\ }\href {\doibase
  10.1103/PhysRevLett.104.125301} {\bibfield  {journal} {\bibinfo  {journal}
  {Phys. Rev. Lett.}\ }\textbf {\bibinfo {volume} {104}},\ \bibinfo {pages}
  {125301} (\bibinfo {year} {2010})}\BibitemShut {NoStop}%
\bibitem [{\citenamefont {Bandyopadhyay}\ \emph {et~al.}(2019)\citenamefont
  {Bandyopadhyay}, \citenamefont {Bai}, \citenamefont {Pal}, \citenamefont
  {Suthar}, \citenamefont {Nath},\ and\ \citenamefont
  {Angom}}]{bandyopadhyay_19}%
  \BibitemOpen
  \bibfield  {author} {\bibinfo {author} {\bibfnamefont {Soumik}\ \bibnamefont
  {Bandyopadhyay}}, \bibinfo {author} {\bibfnamefont {Rukmani}\ \bibnamefont
  {Bai}}, \bibinfo {author} {\bibfnamefont {Sukla}\ \bibnamefont {Pal}},
  \bibinfo {author} {\bibfnamefont {K.}~\bibnamefont {Suthar}}, \bibinfo
  {author} {\bibfnamefont {Rejish}\ \bibnamefont {Nath}}, \ and\ \bibinfo
  {author} {\bibfnamefont {D.}~\bibnamefont {Angom}},\ }\bibfield  {title}
  {\enquote {\bibinfo {title} {Quantum phases of canted dipolar bosons in a
  two-dimensional square optical lattice},}\ }\href {\doibase
  10.1103/PhysRevA.100.053623} {\bibfield  {journal} {\bibinfo  {journal}
  {Phys. Rev. A}\ }\textbf {\bibinfo {volume} {100}},\ \bibinfo {pages}
  {053623} (\bibinfo {year} {2019})}\BibitemShut {NoStop}%
\bibitem [{\citenamefont {Wessel}\ and\ \citenamefont
  {Troyer}(2005)}]{wessel_05}%
  \BibitemOpen
  \bibfield  {author} {\bibinfo {author} {\bibfnamefont {S.}~\bibnamefont
  {Wessel}}\ and\ \bibinfo {author} {\bibfnamefont {M.}~\bibnamefont
  {Troyer}},\ }\bibfield  {title} {\enquote {\bibinfo {title} {Supersolid
  hard-core bosons on the triangular lattice},}\ }\href {\doibase
  10.1103/PhysRevLett.95.127205} {\bibfield  {journal} {\bibinfo  {journal}
  {Phys. Rev. Lett.}\ }\textbf {\bibinfo {volume} {95}},\ \bibinfo {pages}
  {127205} (\bibinfo {year} {2005})}\BibitemShut {NoStop}%
\bibitem [{\citenamefont {Heidarian}\ and\ \citenamefont
  {Damle}(2005)}]{heidarian_05}%
  \BibitemOpen
  \bibfield  {author} {\bibinfo {author} {\bibfnamefont {D.}~\bibnamefont
  {Heidarian}}\ and\ \bibinfo {author} {\bibfnamefont {K.}~\bibnamefont
  {Damle}},\ }\bibfield  {title} {\enquote {\bibinfo {title} {Persistent
  supersolid phase of hard-core bosons on the triangular lattice},}\ }\href
  {\doibase 10.1103/PhysRevLett.95.127206} {\bibfield  {journal} {\bibinfo
  {journal} {Phys. Rev. Lett.}\ }\textbf {\bibinfo {volume} {95}},\ \bibinfo
  {pages} {127206} (\bibinfo {year} {2005})}\BibitemShut {NoStop}%
\bibitem [{\citenamefont {Melko}\ \emph {et~al.}(2005)\citenamefont {Melko},
  \citenamefont {Paramekanti}, \citenamefont {Burkov}, \citenamefont
  {Vishwanath}, \citenamefont {Sheng},\ and\ \citenamefont
  {Balents}}]{melko_05}%
  \BibitemOpen
  \bibfield  {author} {\bibinfo {author} {\bibfnamefont {R.~G.}\ \bibnamefont
  {Melko}}, \bibinfo {author} {\bibfnamefont {A.}~\bibnamefont {Paramekanti}},
  \bibinfo {author} {\bibfnamefont {A.~A.}\ \bibnamefont {Burkov}}, \bibinfo
  {author} {\bibfnamefont {A.}~\bibnamefont {Vishwanath}}, \bibinfo {author}
  {\bibfnamefont {D.~N.}\ \bibnamefont {Sheng}}, \ and\ \bibinfo {author}
  {\bibfnamefont {L.}~\bibnamefont {Balents}},\ }\bibfield  {title} {\enquote
  {\bibinfo {title} {Supersolid order from disorder: Hard-core bosons on the
  triangular lattice},}\ }\href {\doibase 10.1103/PhysRevLett.95.127207}
  {\bibfield  {journal} {\bibinfo  {journal} {Phys. Rev. Lett.}\ }\textbf
  {\bibinfo {volume} {95}},\ \bibinfo {pages} {127207} (\bibinfo {year}
  {2005})}\BibitemShut {NoStop}%
\bibitem [{\citenamefont {Boninsegni}\ and\ \citenamefont
  {Prokof'ev}(2005)}]{boninsegni_05}%
  \BibitemOpen
  \bibfield  {author} {\bibinfo {author} {\bibfnamefont {M.}~\bibnamefont
  {Boninsegni}}\ and\ \bibinfo {author} {\bibfnamefont {N.}~\bibnamefont
  {Prokof'ev}},\ }\bibfield  {title} {\enquote {\bibinfo {title} {Supersolid
  phase of hard-core bosons on a triangular lattice},}\ }\href {\doibase
  10.1103/PhysRevLett.95.237204} {\bibfield  {journal} {\bibinfo  {journal}
  {Phys. Rev. Lett.}\ }\textbf {\bibinfo {volume} {95}},\ \bibinfo {pages}
  {237204} (\bibinfo {year} {2005})}\BibitemShut {NoStop}%
\bibitem [{\citenamefont {Sen}\ \emph {et~al.}(2008)\citenamefont {Sen},
  \citenamefont {Dutt}, \citenamefont {Damle},\ and\ \citenamefont
  {Moessner}}]{sen_08}%
  \BibitemOpen
  \bibfield  {author} {\bibinfo {author} {\bibfnamefont {A.}~\bibnamefont
  {Sen}}, \bibinfo {author} {\bibfnamefont {P.}~\bibnamefont {Dutt}}, \bibinfo
  {author} {\bibfnamefont {K.}~\bibnamefont {Damle}}, \ and\ \bibinfo {author}
  {\bibfnamefont {R.}~\bibnamefont {Moessner}},\ }\bibfield  {title} {\enquote
  {\bibinfo {title} {Variational wave-function study of the triangular lattice
  supersolid},}\ }\href {\doibase 10.1103/PhysRevLett.100.147204} {\bibfield
  {journal} {\bibinfo  {journal} {Phys. Rev. Lett.}\ }\textbf {\bibinfo
  {volume} {100}},\ \bibinfo {pages} {147204} (\bibinfo {year}
  {2008})}\BibitemShut {NoStop}%
\bibitem [{\citenamefont {Yamamoto}\ \emph
  {et~al.}(2012{\natexlab{a}})\citenamefont {Yamamoto}, \citenamefont
  {Danshita},\ and\ \citenamefont {S\'a~de Melo}}]{yamamoto_12}%
  \BibitemOpen
  \bibfield  {author} {\bibinfo {author} {\bibfnamefont {D.}~\bibnamefont
  {Yamamoto}}, \bibinfo {author} {\bibfnamefont {I.}~\bibnamefont {Danshita}},
  \ and\ \bibinfo {author} {\bibfnamefont {C.~A.~R.}\ \bibnamefont {S\'a~de
  Melo}},\ }\bibfield  {title} {\enquote {\bibinfo {title} {Dipolar bosons in
  triangular optical lattices: {Q}uantum phase transitions and anomalous
  hysteresis},}\ }\href {\doibase 10.1103/PhysRevA.85.021601} {\bibfield
  {journal} {\bibinfo  {journal} {Phys. Rev. A}\ }\textbf {\bibinfo {volume}
  {85}},\ \bibinfo {pages} {021601(R)} (\bibinfo {year}
  {2012}{\natexlab{a}})}\BibitemShut {NoStop}%
\bibitem [{\citenamefont {Isakov}\ \emph {et~al.}(2006)\citenamefont {Isakov},
  \citenamefont {Wessel}, \citenamefont {Melko}, \citenamefont {Sengupta},\
  and\ \citenamefont {Kim}}]{isakov_06}%
  \BibitemOpen
  \bibfield  {author} {\bibinfo {author} {\bibfnamefont {S.~V.}\ \bibnamefont
  {Isakov}}, \bibinfo {author} {\bibfnamefont {S.}~\bibnamefont {Wessel}},
  \bibinfo {author} {\bibfnamefont {R.~G.}\ \bibnamefont {Melko}}, \bibinfo
  {author} {\bibfnamefont {K.}~\bibnamefont {Sengupta}}, \ and\ \bibinfo
  {author} {\bibfnamefont {Y.~B.}\ \bibnamefont {Kim}},\ }\bibfield  {title}
  {\enquote {\bibinfo {title} {Hard-core bosons on the kagome lattice:
  {V}alence-bond solids and their quantum melting},}\ }\href {\doibase
  10.1103/PhysRevLett.97.147202} {\bibfield  {journal} {\bibinfo  {journal}
  {Phys. Rev. Lett.}\ }\textbf {\bibinfo {volume} {97}},\ \bibinfo {pages}
  {147202} (\bibinfo {year} {2006})}\BibitemShut {NoStop}%
\bibitem [{\citenamefont {Huerga}\ \emph {et~al.}(2016)\citenamefont {Huerga},
  \citenamefont {Capponi}, \citenamefont {Dukelsky},\ and\ \citenamefont
  {Ortiz}}]{huerga_16}%
  \BibitemOpen
  \bibfield  {author} {\bibinfo {author} {\bibfnamefont {D.}~\bibnamefont
  {Huerga}}, \bibinfo {author} {\bibfnamefont {S.}~\bibnamefont {Capponi}},
  \bibinfo {author} {\bibfnamefont {J.}~\bibnamefont {Dukelsky}}, \ and\
  \bibinfo {author} {\bibfnamefont {G.}~\bibnamefont {Ortiz}},\ }\bibfield
  {title} {\enquote {\bibinfo {title} {Staircase of crystal phases of hard-core
  bosons on the kagome lattice},}\ }\href {\doibase 10.1103/PhysRevB.94.165124}
  {\bibfield  {journal} {\bibinfo  {journal} {Phys. Rev. B}\ }\textbf {\bibinfo
  {volume} {94}},\ \bibinfo {pages} {165124} (\bibinfo {year}
  {2016})}\BibitemShut {NoStop}%
\bibitem [{\citenamefont {Trefzger}\ \emph {et~al.}(2009)\citenamefont
  {Trefzger}, \citenamefont {Menotti},\ and\ \citenamefont
  {Lewenstein}}]{trefzger_09}%
  \BibitemOpen
  \bibfield  {author} {\bibinfo {author} {\bibfnamefont {C.}~\bibnamefont
  {Trefzger}}, \bibinfo {author} {\bibfnamefont {C.}~\bibnamefont {Menotti}}, \
  and\ \bibinfo {author} {\bibfnamefont {M.}~\bibnamefont {Lewenstein}},\
  }\bibfield  {title} {\enquote {\bibinfo {title} {Pair-supersolid phase in a
  bilayer system of dipolar lattice bosons},}\ }\href {\doibase
  10.1103/PhysRevLett.103.035304} {\bibfield  {journal} {\bibinfo  {journal}
  {Phys. Rev. Lett.}\ }\textbf {\bibinfo {volume} {103}},\ \bibinfo {pages}
  {035304} (\bibinfo {year} {2009})}\BibitemShut {NoStop}%
\bibitem [{\citenamefont {Yamamoto}\ \emph {et~al.}(2009)\citenamefont
  {Yamamoto}, \citenamefont {Todo},\ and\ \citenamefont
  {Miyashita}}]{yamamoto_09}%
  \BibitemOpen
  \bibfield  {author} {\bibinfo {author} {\bibfnamefont {K.}~\bibnamefont
  {Yamamoto}}, \bibinfo {author} {\bibfnamefont {S.}~\bibnamefont {Todo}}, \
  and\ \bibinfo {author} {\bibfnamefont {S.}~\bibnamefont {Miyashita}},\
  }\bibfield  {title} {\enquote {\bibinfo {title} {Successive phase transitions
  at finite temperatures toward the supersolid state in a three-dimensional
  extended {B}ose-{H}ubbard model},}\ }\href {\doibase
  10.1103/PhysRevB.79.094503} {\bibfield  {journal} {\bibinfo  {journal} {Phys.
  Rev. B}\ }\textbf {\bibinfo {volume} {79}},\ \bibinfo {pages} {094503}
  (\bibinfo {year} {2009})}\BibitemShut {NoStop}%
\bibitem [{\citenamefont {Xi}\ \emph {et~al.}(2011)\citenamefont {Xi},
  \citenamefont {Ye}, \citenamefont {Chen}, \citenamefont {Zhang},\ and\
  \citenamefont {Su}}]{xi_11}%
  \BibitemOpen
  \bibfield  {author} {\bibinfo {author} {\bibfnamefont {B.}~\bibnamefont
  {Xi}}, \bibinfo {author} {\bibfnamefont {F.}~\bibnamefont {Ye}}, \bibinfo
  {author} {\bibfnamefont {W.}~\bibnamefont {Chen}}, \bibinfo {author}
  {\bibfnamefont {F.}~\bibnamefont {Zhang}}, \ and\ \bibinfo {author}
  {\bibfnamefont {G.}~\bibnamefont {Su}},\ }\bibfield  {title} {\enquote
  {\bibinfo {title} {Global phase diagram of three-dimensional extended boson
  {H}ubbard model: A continuous-time quantum {M}onte {C}arlo study},}\ }\href
  {\doibase 10.1103/PhysRevB.84.054512} {\bibfield  {journal} {\bibinfo
  {journal} {Phys. Rev. B}\ }\textbf {\bibinfo {volume} {84}},\ \bibinfo
  {pages} {054512} (\bibinfo {year} {2011})}\BibitemShut {NoStop}%
\bibitem [{\citenamefont {Ohgoe}\ \emph
  {et~al.}(2012{\natexlab{b}})\citenamefont {Ohgoe}, \citenamefont {Suzuki},\
  and\ \citenamefont {Kawashima}}]{ohgoe_12}%
  \BibitemOpen
  \bibfield  {author} {\bibinfo {author} {\bibfnamefont {T.}~\bibnamefont
  {Ohgoe}}, \bibinfo {author} {\bibfnamefont {T.}~\bibnamefont {Suzuki}}, \
  and\ \bibinfo {author} {\bibfnamefont {N.}~\bibnamefont {Kawashima}},\
  }\bibfield  {title} {\enquote {\bibinfo {title} {Commensurate supersolid of
  three-dimensional lattice bosons},}\ }\href {\doibase
  10.1103/PhysRevLett.108.185302} {\bibfield  {journal} {\bibinfo  {journal}
  {Phys. Rev. Lett.}\ }\textbf {\bibinfo {volume} {108}},\ \bibinfo {pages}
  {185302} (\bibinfo {year} {2012}{\natexlab{b}})}\BibitemShut {NoStop}%
\bibitem [{\citenamefont {Kuno}\ \emph {et~al.}(2017)\citenamefont {Kuno},
  \citenamefont {Shimizu},\ and\ \citenamefont {Ichinose}}]{kuno_17}%
  \BibitemOpen
  \bibfield  {author} {\bibinfo {author} {\bibfnamefont {Y.}~\bibnamefont
  {Kuno}}, \bibinfo {author} {\bibfnamefont {K.}~\bibnamefont {Shimizu}}, \
  and\ \bibinfo {author} {\bibfnamefont {I.}~\bibnamefont {Ichinose}},\
  }\bibfield  {title} {\enquote {\bibinfo {title} {Bosonic analogs of the
  fractional quantum {H}all state in the vicinity of {M}ott states},}\ }\href
  {\doibase 10.1103/PhysRevA.95.013607} {\bibfield  {journal} {\bibinfo
  {journal} {Phys. Rev. A}\ }\textbf {\bibinfo {volume} {95}},\ \bibinfo
  {pages} {013607} (\bibinfo {year} {2017})}\BibitemShut {NoStop}%
\bibitem [{\citenamefont {Kuno}\ \emph {et~al.}(2015)\citenamefont {Kuno},
  \citenamefont {Nakafuji},\ and\ \citenamefont {Ichinose}}]{kuno_15}%
  \BibitemOpen
  \bibfield  {author} {\bibinfo {author} {\bibfnamefont {Y.}~\bibnamefont
  {Kuno}}, \bibinfo {author} {\bibfnamefont {T.}~\bibnamefont {Nakafuji}}, \
  and\ \bibinfo {author} {\bibfnamefont {I.}~\bibnamefont {Ichinose}},\
  }\bibfield  {title} {\enquote {\bibinfo {title} {Phase diagrams of the
  {B}ose-{H}ubbard model and the {H}aldane-{B}ose-{H}ubbard model with complex
  hopping amplitudes},}\ }\href {\doibase 10.1103/PhysRevA.92.063630}
  {\bibfield  {journal} {\bibinfo  {journal} {Phys. Rev. A}\ }\textbf {\bibinfo
  {volume} {92}},\ \bibinfo {pages} {063630} (\bibinfo {year}
  {2015})}\BibitemShut {NoStop}%
\bibitem [{\citenamefont {Pasquiou}\ \emph {et~al.}(2011)\citenamefont
  {Pasquiou}, \citenamefont {Bismut}, \citenamefont {Mar\'echal}, \citenamefont
  {Pedri}, \citenamefont {Vernac}, \citenamefont {Gorceix},\ and\ \citenamefont
  {Laburthe-Tolra}}]{pasquiou_11}%
  \BibitemOpen
  \bibfield  {author} {\bibinfo {author} {\bibfnamefont {B.}~\bibnamefont
  {Pasquiou}}, \bibinfo {author} {\bibfnamefont {G.}~\bibnamefont {Bismut}},
  \bibinfo {author} {\bibfnamefont {E.}~\bibnamefont {Mar\'echal}}, \bibinfo
  {author} {\bibfnamefont {P.}~\bibnamefont {Pedri}}, \bibinfo {author}
  {\bibfnamefont {L.}~\bibnamefont {Vernac}}, \bibinfo {author} {\bibfnamefont
  {O.}~\bibnamefont {Gorceix}}, \ and\ \bibinfo {author} {\bibfnamefont
  {B.}~\bibnamefont {Laburthe-Tolra}},\ }\bibfield  {title} {\enquote {\bibinfo
  {title} {Spin relaxation and band excitation of a dipolar {B}ose-{E}instein
  condensate in 2d optical lattices},}\ }\href {\doibase
  10.1103/PhysRevLett.106.015301} {\bibfield  {journal} {\bibinfo  {journal}
  {Phys. Rev. Lett.}\ }\textbf {\bibinfo {volume} {106}},\ \bibinfo {pages}
  {015301} (\bibinfo {year} {2011})}\BibitemShut {NoStop}%
\bibitem [{\citenamefont {Baier}\ \emph {et~al.}(2016)\citenamefont {Baier},
  \citenamefont {Mark}, \citenamefont {Petter}, \citenamefont {Aikawa},
  \citenamefont {Chomaz}, \citenamefont {Cai}, \citenamefont {Baranov},
  \citenamefont {Zoller},\ and\ \citenamefont {Ferlaino}}]{baier_16}%
  \BibitemOpen
  \bibfield  {author} {\bibinfo {author} {\bibfnamefont {S.}~\bibnamefont
  {Baier}}, \bibinfo {author} {\bibfnamefont {M.~J.}\ \bibnamefont {Mark}},
  \bibinfo {author} {\bibfnamefont {D.}~\bibnamefont {Petter}}, \bibinfo
  {author} {\bibfnamefont {K.}~\bibnamefont {Aikawa}}, \bibinfo {author}
  {\bibfnamefont {L.}~\bibnamefont {Chomaz}}, \bibinfo {author} {\bibfnamefont
  {Z.}~\bibnamefont {Cai}}, \bibinfo {author} {\bibfnamefont {M.}~\bibnamefont
  {Baranov}}, \bibinfo {author} {\bibfnamefont {P.}~\bibnamefont {Zoller}}, \
  and\ \bibinfo {author} {\bibfnamefont {F.}~\bibnamefont {Ferlaino}},\
  }\bibfield  {title} {\enquote {\bibinfo {title} {Extended {B}ose-{H}ubbard
  models with ultracold magnetic atoms},}\ }\href {\doibase
  10.1126/science.aac9812} {\bibfield  {journal} {\bibinfo  {journal}
  {Science}\ }\textbf {\bibinfo {volume} {352}},\ \bibinfo {pages} {201}
  (\bibinfo {year} {2016})}\BibitemShut {NoStop}%
\bibitem [{\citenamefont {Aidelsburger}\ \emph {et~al.}(2011)\citenamefont
  {Aidelsburger}, \citenamefont {Atala}, \citenamefont {Nascimb\`ene},
  \citenamefont {Trotzky}, \citenamefont {Chen},\ and\ \citenamefont
  {Bloch}}]{aidelsburger_11}%
  \BibitemOpen
  \bibfield  {author} {\bibinfo {author} {\bibfnamefont {M.}~\bibnamefont
  {Aidelsburger}}, \bibinfo {author} {\bibfnamefont {M.}~\bibnamefont {Atala}},
  \bibinfo {author} {\bibfnamefont {S.}~\bibnamefont {Nascimb\`ene}}, \bibinfo
  {author} {\bibfnamefont {S.}~\bibnamefont {Trotzky}}, \bibinfo {author}
  {\bibfnamefont {Y.-A.}\ \bibnamefont {Chen}}, \ and\ \bibinfo {author}
  {\bibfnamefont {I.}~\bibnamefont {Bloch}},\ }\bibfield  {title} {\enquote
  {\bibinfo {title} {Experimental realization of strong effective magnetic
  fields in an optical lattice},}\ }\href {\doibase
  10.1103/PhysRevLett.107.255301} {\bibfield  {journal} {\bibinfo  {journal}
  {Phys. Rev. Lett.}\ }\textbf {\bibinfo {volume} {107}},\ \bibinfo {pages}
  {255301} (\bibinfo {year} {2011})}\BibitemShut {NoStop}%
\bibitem [{\citenamefont {Miyake}\ \emph {et~al.}(2013)\citenamefont {Miyake},
  \citenamefont {Siviloglou}, \citenamefont {Kennedy}, \citenamefont {Burton},\
  and\ \citenamefont {Ketterle}}]{miyake_13}%
  \BibitemOpen
  \bibfield  {author} {\bibinfo {author} {\bibfnamefont {H.}~\bibnamefont
  {Miyake}}, \bibinfo {author} {\bibfnamefont {G.~A.}\ \bibnamefont
  {Siviloglou}}, \bibinfo {author} {\bibfnamefont {C.~J.}\ \bibnamefont
  {Kennedy}}, \bibinfo {author} {\bibfnamefont {W.~C.}\ \bibnamefont {Burton}},
  \ and\ \bibinfo {author} {\bibfnamefont {W.}~\bibnamefont {Ketterle}},\
  }\bibfield  {title} {\enquote {\bibinfo {title} {Realizing the {H}arper
  {H}amiltonian with laser-assisted tunneling in optical lattices},}\ }\href
  {\doibase 10.1103/PhysRevLett.111.185302} {\bibfield  {journal} {\bibinfo
  {journal} {Phys. Rev. Lett.}\ }\textbf {\bibinfo {volume} {111}},\ \bibinfo
  {pages} {185302} (\bibinfo {year} {2013})}\BibitemShut {NoStop}%
\bibitem [{\citenamefont {Atala}\ \emph {et~al.}(2014)\citenamefont {Atala},
  \citenamefont {Aidelsburger}, \citenamefont {Lohse}, \citenamefont
  {Barreiro}, \citenamefont {Paredes},\ and\ \citenamefont {Bloch}}]{atala_14}%
  \BibitemOpen
  \bibfield  {author} {\bibinfo {author} {\bibfnamefont {M.}~\bibnamefont
  {Atala}}, \bibinfo {author} {\bibfnamefont {M.}~\bibnamefont {Aidelsburger}},
  \bibinfo {author} {\bibfnamefont {M.}~\bibnamefont {Lohse}}, \bibinfo
  {author} {\bibfnamefont {J.~T.}\ \bibnamefont {Barreiro}}, \bibinfo {author}
  {\bibfnamefont {B.}~\bibnamefont {Paredes}}, \ and\ \bibinfo {author}
  {\bibfnamefont {I.}~\bibnamefont {Bloch}},\ }\bibfield  {title} {\enquote
  {\bibinfo {title} {Observation of chiral currents with ultracold atoms in
  bosonic ladders},}\ }\href {\doibase 10.1038/nphys2998} {\bibfield  {journal}
  {\bibinfo  {journal} {Nature Physics}\ }\textbf {\bibinfo {volume} {10}},\
  \bibinfo {pages} {588} (\bibinfo {year} {2014})}\BibitemShut {NoStop}%
\bibitem [{\citenamefont {Kennedy}\ \emph {et~al.}(2015)\citenamefont
  {Kennedy}, \citenamefont {Burton}, \citenamefont {Chung},\ and\ \citenamefont
  {Ketterle}}]{kennedy_15}%
  \BibitemOpen
  \bibfield  {author} {\bibinfo {author} {\bibfnamefont {C.~J.}\ \bibnamefont
  {Kennedy}}, \bibinfo {author} {\bibfnamefont {W.~C.}\ \bibnamefont {Burton}},
  \bibinfo {author} {\bibfnamefont {W.~C.}\ \bibnamefont {Chung}}, \ and\
  \bibinfo {author} {\bibfnamefont {W.}~\bibnamefont {Ketterle}},\ }\bibfield
  {title} {\enquote {\bibinfo {title} {Observation of {B}ose-{E}instein
  condensation in a strong synthetic magnetic field},}\ }\href {\doibase
  10.1038/nphys3421} {\bibfield  {journal} {\bibinfo  {journal} {Nature
  Physics}\ }\textbf {\bibinfo {volume} {11}},\ \bibinfo {pages} {859}
  (\bibinfo {year} {2015})}\BibitemShut {NoStop}%
\bibitem [{\citenamefont {Rokhsar}\ and\ \citenamefont
  {Kotliar}(1991)}]{rokshar_91}%
  \BibitemOpen
  \bibfield  {author} {\bibinfo {author} {\bibfnamefont {D.~S.}\ \bibnamefont
  {Rokhsar}}\ and\ \bibinfo {author} {\bibfnamefont {B.~G.}\ \bibnamefont
  {Kotliar}},\ }\bibfield  {title} {\enquote {\bibinfo {title} {Gutzwiller
  projection for bosons},}\ }\href {\doibase 10.1103/PhysRevB.44.10328}
  {\bibfield  {journal} {\bibinfo  {journal} {Phys. Rev. B}\ }\textbf {\bibinfo
  {volume} {44}},\ \bibinfo {pages} {10328} (\bibinfo {year}
  {1991})}\BibitemShut {NoStop}%
\bibitem [{\citenamefont {Krauth}\ \emph {et~al.}(1992)\citenamefont {Krauth},
  \citenamefont {Caffarel},\ and\ \citenamefont {Bouchaud}}]{krauth_92}%
  \BibitemOpen
  \bibfield  {author} {\bibinfo {author} {\bibfnamefont {W.}~\bibnamefont
  {Krauth}}, \bibinfo {author} {\bibfnamefont {M.}~\bibnamefont {Caffarel}}, \
  and\ \bibinfo {author} {\bibfnamefont {J.-P.}\ \bibnamefont {Bouchaud}},\
  }\bibfield  {title} {\enquote {\bibinfo {title} {Gutzwiller wave function for
  a model of strongly interacting bosons},}\ }\href {\doibase
  10.1103/PhysRevB.45.3137} {\bibfield  {journal} {\bibinfo  {journal} {Phys.
  Rev. B}\ }\textbf {\bibinfo {volume} {45}},\ \bibinfo {pages} {3137}
  (\bibinfo {year} {1992})}\BibitemShut {NoStop}%
\bibitem [{\citenamefont {Sheshadri}\ \emph {et~al.}(1993)\citenamefont
  {Sheshadri}, \citenamefont {Krishnamurthy}, \citenamefont {Pandit},\ and\
  \citenamefont {Ramakrishnan}}]{sheshadri_93}%
  \BibitemOpen
  \bibfield  {author} {\bibinfo {author} {\bibfnamefont {K.}~\bibnamefont
  {Sheshadri}}, \bibinfo {author} {\bibfnamefont {H.~R.}\ \bibnamefont
  {Krishnamurthy}}, \bibinfo {author} {\bibfnamefont {R.}~\bibnamefont
  {Pandit}}, \ and\ \bibinfo {author} {\bibfnamefont {T.~V.}\ \bibnamefont
  {Ramakrishnan}},\ }\bibfield  {title} {\enquote {\bibinfo {title} {Superfluid
  and insulating phases in an interacting-boson model: Mean-field theory and
  the {RPA}},}\ }\href {\doibase https://doi.org/10.1209/0295-5075/22/4/004}
  {\bibfield  {journal} {\bibinfo  {journal} {EPL}\ }\textbf {\bibinfo {volume}
  {22}},\ \bibinfo {pages} {257} (\bibinfo {year} {1993})}\BibitemShut
  {NoStop}%
\bibitem [{\citenamefont {Bai}\ \emph {et~al.}(2018)\citenamefont {Bai},
  \citenamefont {Bandyopadhyay}, \citenamefont {Pal}, \citenamefont {Suthar},\
  and\ \citenamefont {Angom}}]{rukmani_18}%
  \BibitemOpen
  \bibfield  {author} {\bibinfo {author} {\bibfnamefont {R.}~\bibnamefont
  {Bai}}, \bibinfo {author} {\bibfnamefont {S.}~\bibnamefont {Bandyopadhyay}},
  \bibinfo {author} {\bibfnamefont {S.}~\bibnamefont {Pal}}, \bibinfo {author}
  {\bibfnamefont {K.}~\bibnamefont {Suthar}}, \ and\ \bibinfo {author}
  {\bibfnamefont {D.}~\bibnamefont {Angom}},\ }\bibfield  {title} {\enquote
  {\bibinfo {title} {Bosonic quantum {H}all states in single-layer
  two-dimensional optical lattices},}\ }\href {\doibase
  10.1103/PhysRevA.98.023606} {\bibfield  {journal} {\bibinfo  {journal} {Phys.
  Rev. A}\ }\textbf {\bibinfo {volume} {98}},\ \bibinfo {pages} {023606}
  (\bibinfo {year} {2018})}\BibitemShut {NoStop}%
\bibitem [{\citenamefont {Pal}\ \emph {et~al.}(2019)\citenamefont {Pal},
  \citenamefont {Bai}, \citenamefont {Bandyopadhyay}, \citenamefont {Suthar},\
  and\ \citenamefont {Angom}}]{pal_19}%
  \BibitemOpen
  \bibfield  {author} {\bibinfo {author} {\bibfnamefont {S.}~\bibnamefont
  {Pal}}, \bibinfo {author} {\bibfnamefont {R.}~\bibnamefont {Bai}}, \bibinfo
  {author} {\bibfnamefont {S.}~\bibnamefont {Bandyopadhyay}}, \bibinfo {author}
  {\bibfnamefont {K.}~\bibnamefont {Suthar}}, \ and\ \bibinfo {author}
  {\bibfnamefont {D.}~\bibnamefont {Angom}},\ }\bibfield  {title} {\enquote
  {\bibinfo {title} {Enhancement of the bose glass phase in the presence of an
  artificial gauge field},}\ }\href {\doibase 10.1103/PhysRevA.99.053610}
  {\bibfield  {journal} {\bibinfo  {journal} {Phys. Rev. A}\ }\textbf {\bibinfo
  {volume} {99}},\ \bibinfo {pages} {053610} (\bibinfo {year}
  {2019})}\BibitemShut {NoStop}%
\bibitem [{\citenamefont {Buonsante}\ \emph {et~al.}(2004)\citenamefont
  {Buonsante}, \citenamefont {Penna},\ and\ \citenamefont
  {Vezzani}}]{buonsante_04}%
  \BibitemOpen
  \bibfield  {author} {\bibinfo {author} {\bibfnamefont {P.}~\bibnamefont
  {Buonsante}}, \bibinfo {author} {\bibfnamefont {V.}~\bibnamefont {Penna}}, \
  and\ \bibinfo {author} {\bibfnamefont {A.}~\bibnamefont {Vezzani}},\
  }\bibfield  {title} {\enquote {\bibinfo {title} {Fractional-filling loophole
  insulator domains for ultracold bosons in optical superlattices},}\ }\href
  {\doibase 10.1103/PhysRevA.70.061603} {\bibfield  {journal} {\bibinfo
  {journal} {Phys. Rev. A}\ }\textbf {\bibinfo {volume} {70}},\ \bibinfo
  {pages} {061603(R)} (\bibinfo {year} {2004})}\BibitemShut {NoStop}%
\bibitem [{\citenamefont {Yamamoto}(2009)}]{yamamoto_09a}%
  \BibitemOpen
  \bibfield  {author} {\bibinfo {author} {\bibfnamefont {D.}~\bibnamefont
  {Yamamoto}},\ }\bibfield  {title} {\enquote {\bibinfo {title} {Correlated
  cluster mean-field theory for spin systems},}\ }\href {\doibase
  10.1103/PhysRevB.79.144427} {\bibfield  {journal} {\bibinfo  {journal} {Phys.
  Rev. B}\ }\textbf {\bibinfo {volume} {79}},\ \bibinfo {pages} {144427}
  (\bibinfo {year} {2009})}\BibitemShut {NoStop}%
\bibitem [{\citenamefont {Pisarski}\ \emph {et~al.}(2011)\citenamefont
  {Pisarski}, \citenamefont {Jones},\ and\ \citenamefont
  {Gooding}}]{pisarski_11}%
  \BibitemOpen
  \bibfield  {author} {\bibinfo {author} {\bibfnamefont {P.}~\bibnamefont
  {Pisarski}}, \bibinfo {author} {\bibfnamefont {R.~M.}\ \bibnamefont {Jones}},
  \ and\ \bibinfo {author} {\bibfnamefont {R.~J.}\ \bibnamefont {Gooding}},\
  }\bibfield  {title} {\enquote {\bibinfo {title} {Application of a multisite
  mean-field theory to the disordered {B}ose-{H}ubbard model},}\ }\href
  {\doibase 10.1103/PhysRevA.83.053608} {\bibfield  {journal} {\bibinfo
  {journal} {Phys. Rev. A}\ }\textbf {\bibinfo {volume} {83}},\ \bibinfo
  {pages} {053608} (\bibinfo {year} {2011})}\BibitemShut {NoStop}%
\bibitem [{\citenamefont {McIntosh}\ \emph {et~al.}(2012)\citenamefont
  {McIntosh}, \citenamefont {Pisarski}, \citenamefont {Gooding},\ and\
  \citenamefont {Zaremba}}]{mcintosh_12}%
  \BibitemOpen
  \bibfield  {author} {\bibinfo {author} {\bibfnamefont {T.}~\bibnamefont
  {McIntosh}}, \bibinfo {author} {\bibfnamefont {P.}~\bibnamefont {Pisarski}},
  \bibinfo {author} {\bibfnamefont {R.~J.}\ \bibnamefont {Gooding}}, \ and\
  \bibinfo {author} {\bibfnamefont {E.}~\bibnamefont {Zaremba}},\ }\bibfield
  {title} {\enquote {\bibinfo {title} {Multisite mean-field theory for cold
  bosonic atoms in optical lattices},}\ }\href {\doibase
  10.1103/PhysRevA.86.013623} {\bibfield  {journal} {\bibinfo  {journal} {Phys.
  Rev. A}\ }\textbf {\bibinfo {volume} {86}},\ \bibinfo {pages} {013623}
  (\bibinfo {year} {2012})}\BibitemShut {NoStop}%
\bibitem [{\citenamefont {L\"uhmann}(2013)}]{luhmann_13}%
  \BibitemOpen
  \bibfield  {author} {\bibinfo {author} {\bibfnamefont {D.-S.}\ \bibnamefont
  {L\"uhmann}},\ }\bibfield  {title} {\enquote {\bibinfo {title} {Cluster
  {G}utzwiller method for bosonic lattice systems},}\ }\href {\doibase
  10.1103/PhysRevA.87.043619} {\bibfield  {journal} {\bibinfo  {journal} {Phys.
  Rev. A}\ }\textbf {\bibinfo {volume} {87}},\ \bibinfo {pages} {043619}
  (\bibinfo {year} {2013})}\BibitemShut {NoStop}%
\bibitem [{\citenamefont {Mahmud}\ \emph {et~al.}(2011)\citenamefont {Mahmud},
  \citenamefont {Duchon}, \citenamefont {Kato}, \citenamefont {Kawashima},
  \citenamefont {Scalettar},\ and\ \citenamefont {Trivedi}}]{mahmud_11}%
  \BibitemOpen
  \bibfield  {author} {\bibinfo {author} {\bibfnamefont {K.~W.}\ \bibnamefont
  {Mahmud}}, \bibinfo {author} {\bibfnamefont {E.~N.}\ \bibnamefont {Duchon}},
  \bibinfo {author} {\bibfnamefont {Y.}~\bibnamefont {Kato}}, \bibinfo {author}
  {\bibfnamefont {N.}~\bibnamefont {Kawashima}}, \bibinfo {author}
  {\bibfnamefont {R.~T.}\ \bibnamefont {Scalettar}}, \ and\ \bibinfo {author}
  {\bibfnamefont {N.}~\bibnamefont {Trivedi}},\ }\bibfield  {title} {\enquote
  {\bibinfo {title} {Finite-temperature study of bosons in a two-dimensional
  optical lattice},}\ }\href {\doibase 10.1103/PhysRevB.84.054302} {\bibfield
  {journal} {\bibinfo  {journal} {Phys. Rev. B}\ }\textbf {\bibinfo {volume}
  {84}},\ \bibinfo {pages} {054302} (\bibinfo {year} {2011})}\BibitemShut
  {NoStop}%
\bibitem [{\citenamefont {de~Forges~de Parny}\ \emph
  {et~al.}(2012)\citenamefont {de~Forges~de Parny}, \citenamefont {H\'ebert},
  \citenamefont {Rousseau},\ and\ \citenamefont {Batrouni}}]{parny_12}%
  \BibitemOpen
  \bibfield  {author} {\bibinfo {author} {\bibfnamefont {L.}~\bibnamefont
  {de~Forges~de Parny}}, \bibinfo {author} {\bibfnamefont {F.}~\bibnamefont
  {H\'ebert}}, \bibinfo {author} {\bibfnamefont {V.~G.}\ \bibnamefont
  {Rousseau}}, \ and\ \bibinfo {author} {\bibfnamefont {G.~G.}\ \bibnamefont
  {Batrouni}},\ }\bibfield  {title} {\enquote {\bibinfo {title} {Finite
  temperature phase diagram of spin-1/2 bosons in two-dimensional optical
  lattice},}\ }\href {\doibase 10.1140/epjb/e2012-30055-9} {\bibfield
  {journal} {\bibinfo  {journal} {Eur. Phys. J. B}\ }\textbf {\bibinfo {volume}
  {85}},\ \bibinfo {pages} {169} (\bibinfo {year} {2012})}\BibitemShut
  {NoStop}%
\bibitem [{\citenamefont {Greiner}\ \emph {et~al.}(2002)\citenamefont
  {Greiner}, \citenamefont {Mandel}, \citenamefont {Esslinger}, \citenamefont
  {Hansch},\ and\ \citenamefont {Bloch}}]{greiner_02}%
  \BibitemOpen
  \bibfield  {author} {\bibinfo {author} {\bibfnamefont {M.}~\bibnamefont
  {Greiner}}, \bibinfo {author} {\bibfnamefont {O.}~\bibnamefont {Mandel}},
  \bibinfo {author} {\bibfnamefont {T.}~\bibnamefont {Esslinger}}, \bibinfo
  {author} {\bibfnamefont {T.~W.}\ \bibnamefont {Hansch}}, \ and\ \bibinfo
  {author} {\bibfnamefont {I.}~\bibnamefont {Bloch}},\ }\bibfield  {title}
  {\enquote {\bibinfo {title} {Quantum phase transition from a superfluid to a
  {M}ott insulator in a gas of ultracold atoms},}\ }\href {\doibase
  10.1038/415039a} {\bibfield  {journal} {\bibinfo  {journal} {Nature
  (London)}\ }\textbf {\bibinfo {volume} {415}},\ \bibinfo {pages} {39}
  (\bibinfo {year} {2002})}\BibitemShut {NoStop}%
\bibitem [{\citenamefont {Iskin}(2011)}]{iskin_11}%
  \BibitemOpen
  \bibfield  {author} {\bibinfo {author} {\bibfnamefont {M.}~\bibnamefont
  {Iskin}},\ }\bibfield  {title} {\enquote {\bibinfo {title} {Route to
  supersolidity for the extended {B}ose-{H}ubbard model},}\ }\href {\doibase
  10.1103/PhysRevA.83.051606} {\bibfield  {journal} {\bibinfo  {journal} {Phys.
  Rev. A}\ }\textbf {\bibinfo {volume} {83}},\ \bibinfo {pages} {051606(R)}
  (\bibinfo {year} {2011})}\BibitemShut {NoStop}%
\bibitem [{\citenamefont {Iskin}(2012)}]{iskin_12}%
  \BibitemOpen
  \bibfield  {author} {\bibinfo {author} {\bibfnamefont {M.}~\bibnamefont
  {Iskin}},\ }\bibfield  {title} {\enquote {\bibinfo {title} {Artificial gauge
  fields for the {B}ose-{H}ubbard model on a checkerboard superlattice and
  extended {B}ose-{H}ubbard model},}\ }\href {\doibase
  10.1140/epjb/e2012-20852-5} {\bibfield  {journal} {\bibinfo  {journal} {Eur.
  Phys. J. B}\ }\textbf {\bibinfo {volume} {85}},\ \bibinfo {pages} {76}
  (\bibinfo {year} {2012})}\BibitemShut {NoStop}%
\bibitem [{\citenamefont {Batrouni}\ \emph {et~al.}(1995)\citenamefont
  {Batrouni}, \citenamefont {Scalettar}, \citenamefont {Zimanyi},\ and\
  \citenamefont {Kampf}}]{batrouni_95}%
  \BibitemOpen
  \bibfield  {author} {\bibinfo {author} {\bibfnamefont {G.~G.}\ \bibnamefont
  {Batrouni}}, \bibinfo {author} {\bibfnamefont {R.~T.}\ \bibnamefont
  {Scalettar}}, \bibinfo {author} {\bibfnamefont {G.~T.}\ \bibnamefont
  {Zimanyi}}, \ and\ \bibinfo {author} {\bibfnamefont {A.~P.}\ \bibnamefont
  {Kampf}},\ }\bibfield  {title} {\enquote {\bibinfo {title} {Supersolids in
  the bose-hubbard hamiltonian},}\ }\href {\doibase
  10.1103/PhysRevLett.74.2527} {\bibfield  {journal} {\bibinfo  {journal}
  {Phys. Rev. Lett.}\ }\textbf {\bibinfo {volume} {74}},\ \bibinfo {pages}
  {2527} (\bibinfo {year} {1995})}\BibitemShut {NoStop}%
\bibitem [{\citenamefont {Boada}\ \emph {et~al.}(2010)\citenamefont {Boada},
  \citenamefont {Celi}, \citenamefont {Latorre},\ and\ \citenamefont
  {Pic{\'{o}}}}]{boada_10}%
  \BibitemOpen
  \bibfield  {author} {\bibinfo {author} {\bibfnamefont {O.}~\bibnamefont
  {Boada}}, \bibinfo {author} {\bibfnamefont {A.}~\bibnamefont {Celi}},
  \bibinfo {author} {\bibfnamefont {J.~I.}\ \bibnamefont {Latorre}}, \ and\
  \bibinfo {author} {\bibfnamefont {V}~\bibnamefont {Pic{\'{o}}}},\ }\bibfield
  {title} {\enquote {\bibinfo {title} {Simulation of gauge transformations on
  systems of ultracold atoms},}\ }\href {\doibase
  10.1088/1367-2630/12/11/113055} {\bibfield  {journal} {\bibinfo  {journal}
  {New Journal of Physics}\ }\textbf {\bibinfo {volume} {12}},\ \bibinfo
  {pages} {113055} (\bibinfo {year} {2010})}\BibitemShut {NoStop}%
\bibitem [{\citenamefont {M\"oller}\ and\ \citenamefont
  {Cooper}(2010)}]{moller_10}%
  \BibitemOpen
  \bibfield  {author} {\bibinfo {author} {\bibfnamefont {G.}~\bibnamefont
  {M\"oller}}\ and\ \bibinfo {author} {\bibfnamefont {N.~R.}\ \bibnamefont
  {Cooper}},\ }\bibfield  {title} {\enquote {\bibinfo {title} {Condensed ground
  states of frustrated {B}ose-{H}ubbard models},}\ }\href {\doibase
  10.1103/PhysRevA.82.063625} {\bibfield  {journal} {\bibinfo  {journal} {Phys.
  Rev. A}\ }\textbf {\bibinfo {volume} {82}},\ \bibinfo {pages} {063625}
  (\bibinfo {year} {2010})}\BibitemShut {NoStop}%
\bibitem [{\citenamefont {Yamamoto}\ \emph
  {et~al.}(2012{\natexlab{b}})\citenamefont {Yamamoto}, \citenamefont
  {Masaki},\ and\ \citenamefont {Danshita}}]{yamamoto_12a}%
  \BibitemOpen
  \bibfield  {author} {\bibinfo {author} {\bibfnamefont {D.}~\bibnamefont
  {Yamamoto}}, \bibinfo {author} {\bibfnamefont {A.}~\bibnamefont {Masaki}}, \
  and\ \bibinfo {author} {\bibfnamefont {I.}~\bibnamefont {Danshita}},\
  }\bibfield  {title} {\enquote {\bibinfo {title} {Quantum phases of hardcore
  bosons with long-range interactions on a square lattice},}\ }\href {\doibase
  10.1103/PhysRevB.86.054516} {\bibfield  {journal} {\bibinfo  {journal} {Phys.
  Rev. B}\ }\textbf {\bibinfo {volume} {86}},\ \bibinfo {pages} {054516}
  (\bibinfo {year} {2012}{\natexlab{b}})}\BibitemShut {NoStop}%
\bibitem [{\citenamefont {Trefzger}\ \emph {et~al.}(2008)\citenamefont
  {Trefzger}, \citenamefont {Menotti},\ and\ \citenamefont
  {Lewenstein}}]{trefzger_08}%
  \BibitemOpen
  \bibfield  {author} {\bibinfo {author} {\bibfnamefont {C.}~\bibnamefont
  {Trefzger}}, \bibinfo {author} {\bibfnamefont {C.}~\bibnamefont {Menotti}}, \
  and\ \bibinfo {author} {\bibfnamefont {M.}~\bibnamefont {Lewenstein}},\
  }\bibfield  {title} {\enquote {\bibinfo {title} {Ultracold dipolar gas in an
  optical lattice: {T}he fate of metastable states},}\ }\href {\doibase
  10.1103/PhysRevA.78.043604} {\bibfield  {journal} {\bibinfo  {journal} {Phys.
  Rev. A}\ }\textbf {\bibinfo {volume} {78}},\ \bibinfo {pages} {043604}
  (\bibinfo {year} {2008})}\BibitemShut {NoStop}%
\bibitem [{\citenamefont {Werner}\ \emph {et~al.}(2005)\citenamefont {Werner},
  \citenamefont {Griesmaier}, \citenamefont {Hensler}, \citenamefont {Stuhler},
  \citenamefont {Pfau}, \citenamefont {Simoni},\ and\ \citenamefont
  {Tiesinga}}]{werner_05}%
  \BibitemOpen
  \bibfield  {author} {\bibinfo {author} {\bibfnamefont {J.}~\bibnamefont
  {Werner}}, \bibinfo {author} {\bibfnamefont {A.}~\bibnamefont {Griesmaier}},
  \bibinfo {author} {\bibfnamefont {S.}~\bibnamefont {Hensler}}, \bibinfo
  {author} {\bibfnamefont {J.}~\bibnamefont {Stuhler}}, \bibinfo {author}
  {\bibfnamefont {T.}~\bibnamefont {Pfau}}, \bibinfo {author} {\bibfnamefont
  {A.}~\bibnamefont {Simoni}}, \ and\ \bibinfo {author} {\bibfnamefont
  {E.}~\bibnamefont {Tiesinga}},\ }\bibfield  {title} {\enquote {\bibinfo
  {title} {Observation of {F}eshbach resonances in an ultracold gas of $^{52}$
  {C}r},}\ }\href {\doibase 10.1103/PhysRevLett.94.183201} {\bibfield
  {journal} {\bibinfo  {journal} {Phys. Rev. Lett.}\ }\textbf {\bibinfo
  {volume} {94}},\ \bibinfo {pages} {183201} (\bibinfo {year}
  {2005})}\BibitemShut {NoStop}%
\bibitem [{\citenamefont {Frisch}\ \emph {et~al.}(2014)\citenamefont {Frisch},
  \citenamefont {Mark}, \citenamefont {Aikawa}, \citenamefont {Ferlaino},
  \citenamefont {Bohn}, \citenamefont {Makrides}, \citenamefont {Petrov},\ and\
  \citenamefont {Kotochigova}}]{frisch_14}%
  \BibitemOpen
  \bibfield  {author} {\bibinfo {author} {\bibfnamefont {A.}~\bibnamefont
  {Frisch}}, \bibinfo {author} {\bibfnamefont {M.}~\bibnamefont {Mark}},
  \bibinfo {author} {\bibfnamefont {K.}~\bibnamefont {Aikawa}}, \bibinfo
  {author} {\bibfnamefont {F.}~\bibnamefont {Ferlaino}}, \bibinfo {author}
  {\bibfnamefont {J.~L.}\ \bibnamefont {Bohn}}, \bibinfo {author}
  {\bibfnamefont {C.}~\bibnamefont {Makrides}}, \bibinfo {author}
  {\bibfnamefont {A.}~\bibnamefont {Petrov}}, \ and\ \bibinfo {author}
  {\bibfnamefont {S.}~\bibnamefont {Kotochigova}},\ }\bibfield  {title}
  {\enquote {\bibinfo {title} {Quantum chaos in ultracold collisions of
  gas-phase erbium atoms},}\ }\href {\doibase 10.1038/nature13137} {\bibfield
  {journal} {\bibinfo  {journal} {Nature (London)}\ }\textbf {\bibinfo {volume}
  {507}},\ \bibinfo {pages} {475} (\bibinfo {year} {2014})}\BibitemShut
  {NoStop}%
\bibitem [{\citenamefont {Maier}\ \emph {et~al.}(2015)\citenamefont {Maier},
  \citenamefont {Ferrier-Barbut}, \citenamefont {Kadau}, \citenamefont
  {Schmitt}, \citenamefont {Wenzel}, \citenamefont {Wink}, \citenamefont
  {Pfau}, \citenamefont {Jachymski},\ and\ \citenamefont
  {Julienne}}]{maier_15}%
  \BibitemOpen
  \bibfield  {author} {\bibinfo {author} {\bibfnamefont {T.}~\bibnamefont
  {Maier}}, \bibinfo {author} {\bibfnamefont {I.}~\bibnamefont
  {Ferrier-Barbut}}, \bibinfo {author} {\bibfnamefont {H.}~\bibnamefont
  {Kadau}}, \bibinfo {author} {\bibfnamefont {M.}~\bibnamefont {Schmitt}},
  \bibinfo {author} {\bibfnamefont {M.}~\bibnamefont {Wenzel}}, \bibinfo
  {author} {\bibfnamefont {C.}~\bibnamefont {Wink}}, \bibinfo {author}
  {\bibfnamefont {T.}~\bibnamefont {Pfau}}, \bibinfo {author} {\bibfnamefont
  {K.}~\bibnamefont {Jachymski}}, \ and\ \bibinfo {author} {\bibfnamefont
  {P.~S.}\ \bibnamefont {Julienne}},\ }\bibfield  {title} {\enquote {\bibinfo
  {title} {Broad universal {F}eshbach resonances in the chaotic spectrum of
  {d}ysprosium atoms},}\ }\href {\doibase 10.1103/PhysRevA.92.060702}
  {\bibfield  {journal} {\bibinfo  {journal} {Phys. Rev. A}\ }\textbf {\bibinfo
  {volume} {92}},\ \bibinfo {pages} {060702(R)} (\bibinfo {year}
  {2015})}\BibitemShut {NoStop}%
\bibitem [{\citenamefont {Lucioni}\ \emph {et~al.}(2018)\citenamefont
  {Lucioni}, \citenamefont {Tanzi}, \citenamefont {Fregosi}, \citenamefont
  {Catani}, \citenamefont {Gozzini}, \citenamefont {Inguscio}, \citenamefont
  {Fioretti}, \citenamefont {Gabbanini},\ and\ \citenamefont
  {Modugno}}]{lucioni_18}%
  \BibitemOpen
  \bibfield  {author} {\bibinfo {author} {\bibfnamefont {E.}~\bibnamefont
  {Lucioni}}, \bibinfo {author} {\bibfnamefont {L.}~\bibnamefont {Tanzi}},
  \bibinfo {author} {\bibfnamefont {A.}~\bibnamefont {Fregosi}}, \bibinfo
  {author} {\bibfnamefont {J.}~\bibnamefont {Catani}}, \bibinfo {author}
  {\bibfnamefont {S.}~\bibnamefont {Gozzini}}, \bibinfo {author} {\bibfnamefont
  {M.}~\bibnamefont {Inguscio}}, \bibinfo {author} {\bibfnamefont
  {A.}~\bibnamefont {Fioretti}}, \bibinfo {author} {\bibfnamefont
  {C.}~\bibnamefont {Gabbanini}}, \ and\ \bibinfo {author} {\bibfnamefont
  {G.}~\bibnamefont {Modugno}},\ }\bibfield  {title} {\enquote {\bibinfo
  {title} {Dysprosium dipolar {B}ose-{E}instein condensate with broad
  {F}eshbach resonances},}\ }\href {\doibase 10.1103/PhysRevA.97.060701}
  {\bibfield  {journal} {\bibinfo  {journal} {Phys. Rev. A}\ }\textbf {\bibinfo
  {volume} {97}},\ \bibinfo {pages} {060701(R)} (\bibinfo {year}
  {2018})}\BibitemShut {NoStop}%
\end{thebibliography}%

\end{document}